\documentclass[11pt,twoside]{article}

\usepackage{xcolor}

\usepackage{fullpage}

\usepackage{epsf}
\usepackage{fancyhdr}
\usepackage{graphics}
\usepackage{graphicx}
\usepackage{psfrag}
\usepackage{comment}

\usepackage[linesnumbered,ruled]{algorithm2e}% http://ctan.org/pkg/algorithm2e
\DontPrintSemicolon	

\usepackage{color}
\usepackage{amsthm}
\usepackage{amsfonts}
\usepackage{amsmath}
\usepackage{bm}
\usepackage{amssymb,bbm}
\usepackage{natbib}
\usepackage{algorithmic}
\usepackage{url}
\usepackage[colorlinks=True,linkcolor=magenta,citecolor=blue,urlcolor=blue,pagebackref=true,backref=true]
{hyperref}
\renewcommand*{\backref}[1]{\ifx#1\relax \else Page #1 \fi}
\renewcommand*{\backrefalt}[4]{%
    \ifcase #1 \footnotesize{(Not cited.)}%
    \or        \footnotesize{(Cited on page~#2.)}%
    \else      \footnotesize{(Cited on pages~#2.)}%
    \fi}

\usepackage{nicefrac}

\def\half{\hbox{$1\over2$}}

\def\Si{\mbox{Si}}
\def\hatt{\widehat}
\def\N{\mbox{N}}

\usepackage{chngpage}

 \usepackage{tabularx}%

\usepackage{enumitem}
\usepackage{booktabs}
\usepackage{pbox}

\usepackage{caption,subcaption}

\usepackage{mathtools}

\usepackage{fullpage}

\newcommand{\brackets}[1]{\left[ #1 \right]}
\newcommand{\parenth}[1]{\left( #1 \right)}

\newcommand{\abss}[1]{\left| #1 \right |}

\newcommand{\radius}{\ensuremath{R}}

\newtheoremstyle{named}{}{}{\itshape}{}{\bfseries}{.}{.5em}{\thmnote{#3's }#1}
\theoremstyle{named}

\theoremstyle{plain}

\newtheorem{theorem}{Theorem}

\newtheorem{definition}{Definition}

\newlength{\widebarargwidth}
\newlength{\widebarargheight}
\newlength{\widebarargdepth}

\makeatletter
\long\def\@makecaption#1#2{
        \vskip 0.8ex
        \setbox\@tempboxa\hbox{\small {\bf #1:} #2}
        \parindent 1.5em  %% How can we use the global value of this???
        \dimen0=\hsize
        \advance\dimen0 by -3em
        \ifdim \wd\@tempboxa >\dimen0
                \hbox to \hsize{
                        \parindent 0em
                        \hfil
                        \parbox{\dimen0}{\def\baselinestretch{0.96}\small
                                {\bf #1.} #2
                                }
                        \hfil}
        \else \hbox to \hsize{\hfil \box\@tempboxa \hfil}
        \fi
        }
\makeatother

\long\def\comment#1{}
\definecolor{battleshipgrey}{rgb}{0.52, 0.52, 0.51}
\definecolor{darkgray}{rgb}{0.66, 0.66, 0.66}
\definecolor{darkgreen}{rgb}{0.0, 0.2, 0.13}
\definecolor{darkspringgreen}{rgb}{0.09, 0.45, 0.27}
\definecolor{dukeblue}{rgb}{0.0, 0.0, 0.61}
\definecolor{olivedrab7}{rgb}{0.24, 0.2, 0.12}
\definecolor{darkblue}{rgb}{0.0, 0.0, 0.55}
\definecolor{darkscarlet}{rgb}{0.34, 0.01, 0.1}
\definecolor{candyapplered}{rgb}{1.0, 0.03, 0.0}
\definecolor{ao(english)}{rgb}{0.0, 0.5, 0.0}
\definecolor{applegreen}{rgb}{0.55, 0.71, 0.0}

\begin{document}
\begin{center}

{\bf{\LARGE{Statistical Analysis from the Fourier Integral Theorem}}}
  
\vspace*{.2in}
{\large{
\begin{tabular}{cc}
Nhat Ho$^{\diamond}$ & Stephen G. Walker$^{\diamond, \flat}$ \\
\end{tabular}
}}

\vspace*{.2in}

\begin{tabular}{c}
Department of Statistics and Data Sciences, University of Texas at Austin$^\diamond$, \\
Department of Mathematics, University of Texas at Austin$^\flat$ \\
\end{tabular}

\today

\vspace*{.2in}

\begin{abstract}
Taking the Fourier integral theorem as our starting point, in this paper we focus on natural Monte Carlo and fully nonparametric estimators of multivariate distributions and conditional distribution functions. We do this without the need for any estimated covariance matrix or dependence structure between variables. These aspects arise immediately from the integral theorem.  Being able to model multivariate data sets using conditional distribution functions we can study a number of problems, such as prediction for Markov processes, estimation of mixing distribution functions which depend on covariates, and general multivariate data. Estimators are explicit Monte Carlo based and require no recursive or iterative algorithms.
\end{abstract}
\end{center}


\section{Introduction}
Estimation of a multivariate distribution or conditional distribution function necessarily requires the construction of a dependence structure between variables. This is usually applied to a multivariate density function and then integrated to obtain the corresponding distribution function; see, for example, \cite{Jin99}. 
For example, for a bivariate set of observations, marginal distributions are easy to estimate and can be achieved either parametrically or nonparametrically; the latter including empirical distributions or smoothing estimators.  On the other hand, constructing a dependent model is more problematic, particulary from a nonparametric perspective. See, for example, \cite{Panaretos12}, \cite{Wand92}, \cite{WJones93}, \cite{Stanis93}, and \cite{Chacon18}. 

The most common approach is to use smoothing methods based on kernels, such as the Gaussian kernel. However, this would need the estimation of a covariance matrix. Even if the interest focuses on a conditional distribution function with multiple conditioning variables, the same problem arises as with the multivariate kernel methods. As a consequence, a number of authors consider only a single covariate, such as \cite{Hall99} and \cite{Gijbels14}.

The estimation of a dependence structure can be achieved using a nonparametric estimation of a copula function. See, for example, \cite{Chen2007} and \cite{Geenens2017}. However, this is far from trivial and is by no means a popular or common approach. Even with copulas, kernel methods are employed and now with the additional burden of dealing with boundary values.
%However, kernel methods are still used and now with the additional concern of boundary values.

We focus on distribution functions with multiple conditional variables, and we manage this fully nonparametrically without the need for the construction of a full covariance bandwidth matrix.
Hence, the aim in this paper is to show how it is possible to work with multivariate conditional distribution functions and obtain nonparametric estimators while avoiding the need to estimate any dependence structure.  We also consider a variety of other nonparametric estimation of multivariate functions which are connected to the distribution functions, such as quantile functions.

The starting point is the Fourier Integral Theorem. Briefly here, for a suitable function $m$ on $\mathbb{R}^d$, the theorem is given by
\begin{equation} \label{fit}
m(y)  =\lim_{R\to\infty }\frac{1}{\pi^d}\int_{\mathbb{R}^{d}} \int_{[0,R]^{d}}\prod_{j=1}^d \cos(s_j(y_j-x_j))\,m(x)\,ds\, dx,
\end{equation}
for any $y \in \mathbb{R}^{d}$ where each of the two integrals listed are $d$--fold. In \cite{Ho21}, the authors consider multivariate density and multivariate regression estimation using equation (\ref{fit}). Natural nonparametric Monte Carlo estimators are obtained. One drawback, despite excellent convergence properties, is that density function estimators are not guaranteed to 
be densities. However, it is easy to use equation (\ref{fit}) to obtain natural Monte Carlo estimators of distribution functions and these can easily be adapted, using for example, isotonic regression to be proper distribution functions.
These are easily sampled, with an unlimited amount of samples available, which can then be used to undertake statistical inference via a ``generative model'' approach.
 
The important feature of equation~(\ref{fit}) which forms the basis of the paper is that whatever dependence structure exists within $m(x)$, it is transferred to $m(y)$ via the integration and the use of \emph{independent} terms of the type
$$\prod_{j=1}^d \frac{\sin(R(y_j-x_j))}{y_j-x_j},$$
for some suitable choice of $R>0$.
This is quite remarkable and as we shall see allows us to obtain nonparametric Monte Carlo estimators of, for example, distribution functions without requiring the construction of any dependence structure.

A number of authors have used the Fourier kernel for density estimation, see \cite{Parzen62} and \cite{Davis75}. 
However, there has not to our knowledge been any attempt to use the kernel for multivariate density estimation. The reason could well be the apparent inability to model a covariance matrix within the kernel. On the other hand, from the foundations of the sin kernel arising from the Fourier integral theorem, it is clear that a covariance matrix is not required.  

We will be focusing on  distribution functions; hence adapting equation~(\ref{fit}) to this scenario, and for ease of introduction, we first restrict the setting to one dimension. Further, we exchange the limit with $R$ to some fixed value, to get
\begin{equation}\label{dist}
F_R(y)=\half+\frac{1}{\pi}\int \Si(R(y-x))\,f(x)\,d x,
\end{equation}
where $\Si(z)=\int_0^z\sin (x)/x\,dx$.
Equation (\ref{dist}) comes from equation (\ref{fit}) using an integration by parts along with the fact that $\int_0^\infty (\sin x)/x\,d x=\pi/2$.
Here $f$ is a density function on $\mathbb{R}$, $F_R(y)$ the approximated distribution function, with $F_R(y)\to F(y)$ as $R\to \infty$, and
%The key idea now is that the distribution can be estimated from a sample 
%$Y_1,\ldots,Y_n$ via a Monte Carlo estimator
\begin{equation}\label{Fhat}
\widehat{F}_R(y)=\half +\frac{1}{n\pi}\sum_{i=1}^n \Si(R(y-Y_i))
\end{equation}
%is how we would estimate $F_R(y)$ using Monte Carlo methods and a sample $(Y_1,\ldots,Y_n)$.
%As $R\to\infty$, equation~(\ref{Fhat}) converges to the empirical distribution function.
%
%is how we would estimate (\ref{dist}) using Monte Carlo methods from a sample $(Y_1,\ldots,Y_n)$.
%As $R\to\infty$, equation (\ref{Fhat}) converges to the empirical distribution function.
is how we would estimate (\ref{dist}) from a sample $(Y_1,\ldots,Y_n)$. There are some practical aspects to using equation (\ref{Fhat}) to estimate the distribution function, the main one being isotonic regression, if required, to ensure one obtains a proper distribution function. We will also be using equation (\ref{Fhat}) to generate samples when implementing  a generative model.

The important contribution of the paper is that we can estimate conditional distributions with more than one conditioning variable. If using alternative approaches, such as Gaussian kernels, there is a need to install a covariance matrix.  We do  not need to do this so we are estimating conditional distributions fully nonparametrically.
Specific cases we consider include estimating missing outcomes in a Markov process, requiring two conditional variables, and also estimating a conditional mixing distribution.

The layout of the paper is as follows. In Section~\ref{Sec:property_Fourier} we set down the theory for the estimators from the Fourier integral theorem used throughout the paper. We also hint at a class of function for which an integral theorem will hold, including the possibility of a Haar wavelet integral theorem. In Section~\ref{sec:conditiona_distribution} we illustrate with conditional distribution functions; in Section~\ref{Sec:Quantile_Markov} with conditional quantile functions and Markov constructed conditional distributions. Section~\ref{sec:mixing} considers conditional mixing distribution functions. Finally, Section~\ref{sec:discussion} contains a brief discussions and the Appendix contains the proof to the main results.
\section{Properties of the Fourier Integral Theorem}
\label{Sec:property_Fourier}
Before going into the details of applications of the Fourier integral theorem to estimate distribution functions and other related problems, we reconsider the approximation property of the Fourier integral theorem. In the previous work, \cite{Ho21} utilize the tails of the Fourier transform of the function $m(\cdot)$ to characterize the approximation error of the Fourier integral theorem when truncating one of the integrals. However, the proof technique in that work is inherently based on the nice property of sin function in the Fourier integral theorem and is non-trivial to extend to other choices of useful cyclic functions; an example of such a function is in a remark after Theorem~\ref{theorem:approximation_Fourier}. 

In this work, we provide insight into the approximation error between $F(y)$ and $F_R(y)$ via the Riemann sum approximation theorem. This insight can be generalized into any cyclic function which integrates to 0 over the cyclic interval, thereby enriching the family of integral theorems beyond the Fourier integral theorem. Such extensions would provide for example a Haar wavelet integral theorem.

To simplify the presentation, we define
\begin{align}
    m_{R}(y) & : = \frac{1}{\pi^d}\int_{\mathbb{R}^{d}} \int_{[0,R]^{d}} \prod_{j=1}^d \cos(s_j(y_j-x_j))\,m(x)\,ds\, dx \nonumber \\
    & =  \frac{1}{\pi^d}\int_{\mathbb{R}^{d}} \prod_{j = 1}^{d} \frac{\sin(R(y_j-x_j))}{(y_{j} - x_{j})} m(x) \, dx. \label{eq:Fourier_approx}
\end{align}
We start with the following definition of the class of univariate functions that we use throughout our study.

\vspace{0.1in}
\noindent
\begin{definition}
\label{def:class_function}
The univariate function $f(\cdot)$ is said to belong to the class $\mathcal{T}^{K}(\mathbb{R})$ if for any $y \in \mathbb{R}$, the function $g(x) = (f(x) - f(y))/(x - y)$ satisfies the following conditions:
\begin{enumerate}
    \item The function $g$ is differentiable, uniformly continuous up to the $K$-th order, and the limits $\lim_{|x| \to +\infty} | g^{(k)}(x)| = 0$ for any $0 \leq k \leq K$ where $g^{(k)}(.)$ denotes the $k$-th order derivative of $g$;
    \item The integrals $\int_{\mathbb{R}} |g^{(k)}(x)| dx$ are finite  for all $0 \leq k \leq K$.
\end{enumerate}
\end{definition}

\vspace{0.1in}
\noindent
Note that, for the function $g$ in Definition~\ref{def:class_function}, for any $y \in \mathbb{R}$ when $x = y$, we choose $g(y) = f^{(1)}(y)$. An example of function that satisfies Definition~\ref{def:class_function} is Beta density function. Based on Definition~\ref{def:class_function}, we now state the following result.
\begin{theorem}
\label{theorem:approximation_Fourier}
Assume that the univariate functions $m_{j} \in \mathcal{T}^{K_{j}}(\mathbb{R})$ for any $1 \leq j \leq d$ where $K_{1}, \ldots, K_{d}$ are given positive integer numbers. Then, if we have $m(x) = \prod_{j = 1}^{d} m_{j}(x_{j})$ or $m(x) = \sum_{j = 1}^{d} m_{j}(x_{j})$ for any $x = (x_{1}, \ldots, x_{d})$, there exist universal constant $C$ and $\bar{C}$ depending on $d$ such that as long as $R \geq C$ we obtain
\begin{align*}
    \abss{m_{R}(y) - m(y)} \leq \bar{C}/R^{K},
\end{align*}
where $K = \min_{1 \leq j \leq d} \{K_{j}\}$.
\end{theorem}
 \begin{figure}[t]
\centering
\begin{subfigure}[t]{0.48\textwidth}
\includegraphics[width=1\textwidth]{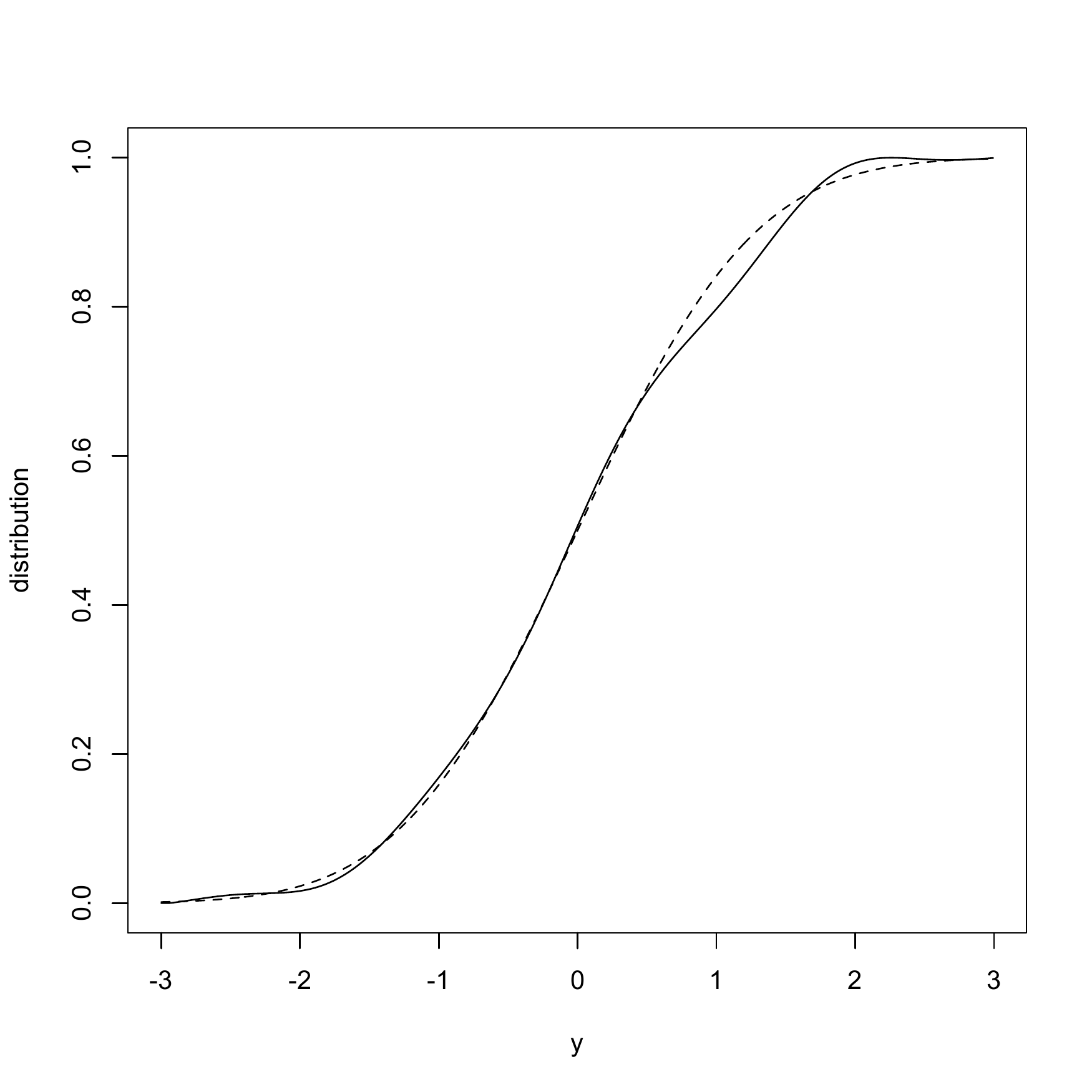}
\caption{}
\end{subfigure}
\begin{subfigure}[t]{0.48\textwidth}
\includegraphics[width=1\textwidth]{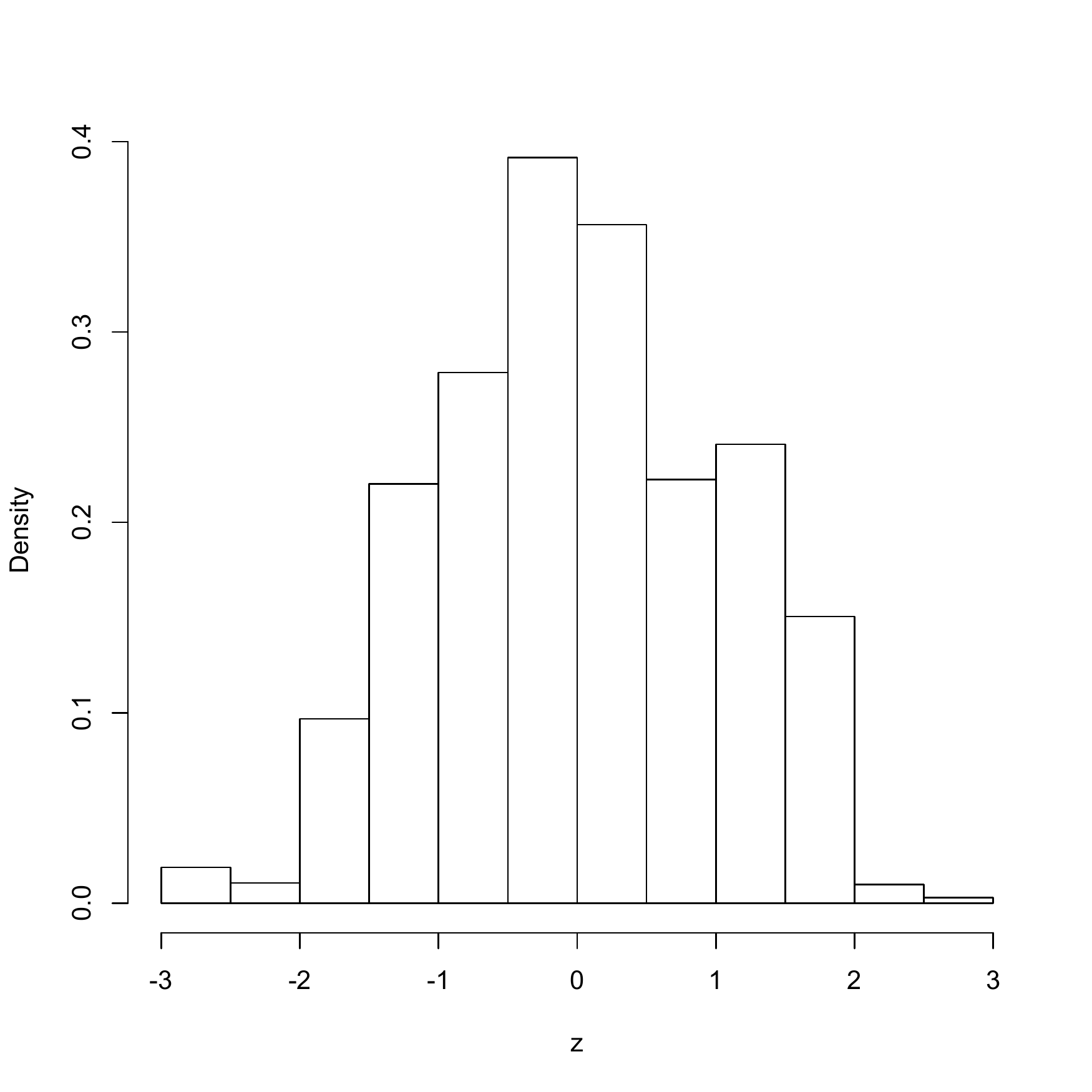}
\caption{}
\end{subfigure}
\caption{
Simulation with the cumulative distribution function via the Fourier integral theorem in equations~\eqref{minmax} and~\eqref{sample}. (a) Estimated $\hatt{F}_R$ (bold line) and true distribution (dashed line); (b) Histogram of 10000 samples taken from $\hatt{F}_R$.
}
\label{fig:cumulative_distribution}
\end{figure}
\vspace{0.2in}
\noindent
The proof is presented in the Appendix. To appreciate the proof we demonstrate the key idea in the one dimensional case. Here
$$m_R(y)-m(y)=\frac{1}{\pi}\int_{-\infty}^{+\infty} \frac{\sin(R(y-x))}{y-x}\,(m(x)-m(y))\,d x$$
which we write as 
$$m_R(y)-m(y)=\frac{1}{\pi}\int_{-\infty}^{+\infty} \sin(R(x-y))\,g(x)\,d x,$$
where $g(x)=(m(x)-m(y))/(x-y)$. Without loss of generality set $y=0$ to get
$$m_R(y)-m(y)=\frac{1}{\pi}\int_{-\infty}^{+\infty} \sin(z)\,\epsilon g(z\epsilon)\,d z,$$
where $\epsilon=1/R$. Now due to the cyclic behaviour of the sin function we can write this as
$$m_R(y)-m(y)=\frac{1}{\pi}\int_{0}^{2\pi} \sin (t)\,\sum_{k=-\infty}^{+\infty} \epsilon g(\epsilon(t+2\pi k))\,\,d t.$$
The term
$\sum_{k=-\infty}^{+\infty} \epsilon g(\epsilon(t+2\pi k))$
is a Riemann sum approximation to an integral which converges to a constant, for all $t$, as $\epsilon \to 0$.
The overall convergence to 0 is then a consequence of $\int_0^{2\pi} \sin\,t \,d t=0$. Hence, it is how the Riemann sum converges to a constant which determines the speed at which $m_R(y)-m(y)\to 0$.

\vspace{0.5 em}
\noindent
\textit{Remark.} It is now interesting to note that the sin function here could be replaced by any cyclic function which integrates to 0 over the cyclic interval. For example,
$$\phi(x)=\left\{\begin{array}{ll}
\frac{2}{\pi}(x-2m\pi), & (2m-1/2)\pi<x<(2m+1/2)\pi \\ \\
\frac{2}{\pi}\left[(2m+1)\pi-x\right], & (2m+1/2)\pi<x<(2m+3/2)\pi.
\end{array}\right.
$$
This yields a wavelet integral theorem based on the Haar wavelet. There are potentially many other such functions such as $\phi(x)$ and $\sin x$, and this line of research will be studied in the future.
\begin{figure}[!t]
\includegraphics[width=14cm,height=9cm]{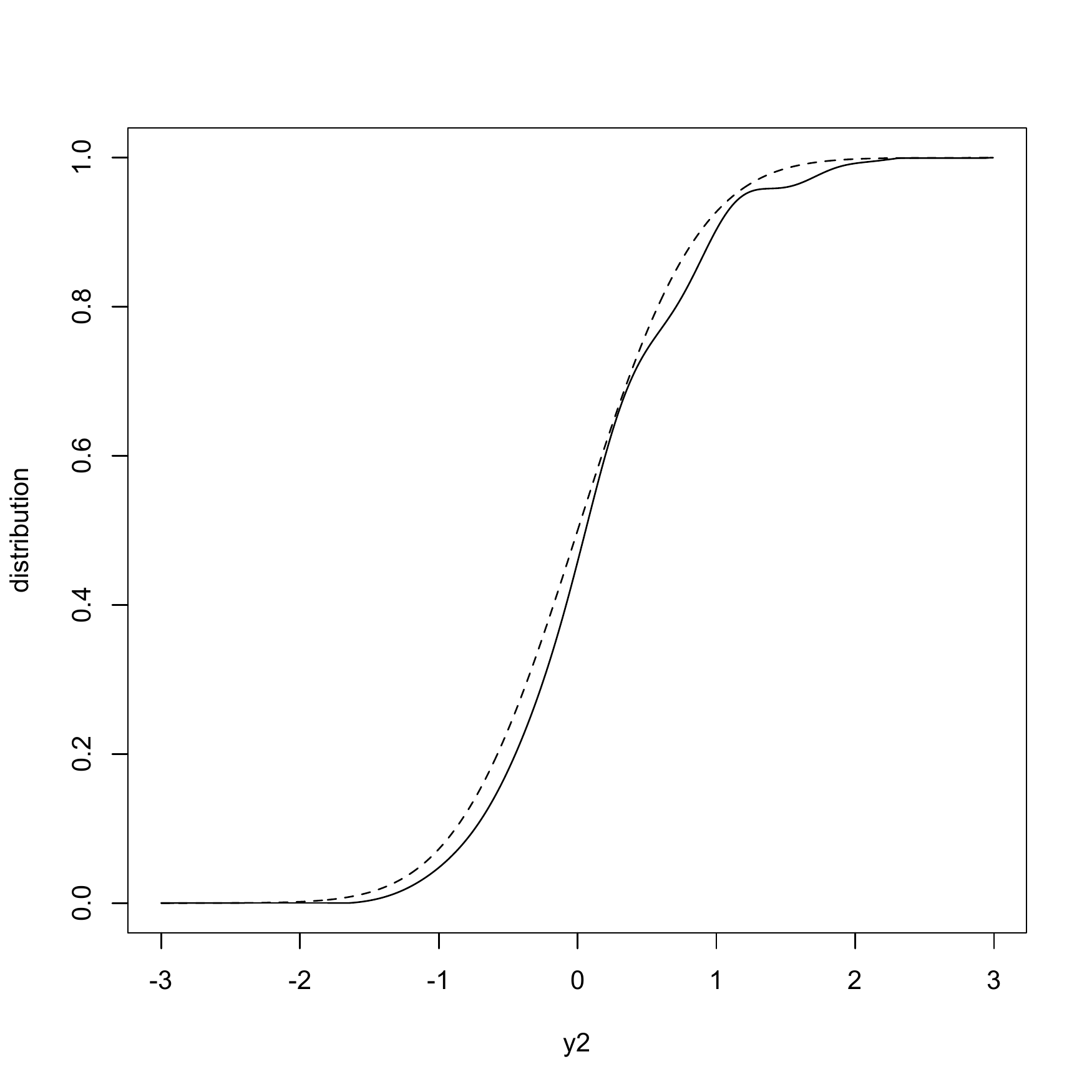}
\caption{Simulation with the conditional distribution function via the Fourier integral theorem. In this figure, we plot the estimated conditional distribution of $y_2$ given $y_1=y_2=0$ (bold line) and the true distribution (dashed line).}
\label{fig3}
\end{figure}
\section{Conditional distribution function}
\label{sec:conditiona_distribution}
The distribution estimator for one dimensional independent and identically distributed (i.i.d.) data is given in equation~(\ref{Fhat}). In order to ensure that $\widehat{F}_R(y)$ lies between 0 and 1 we adapt to
\begin{equation}\label{minmax}
\widehat{F}_R(y)=\min\left\{1,\max\left\{0,\widehat{F}_R(y)\right\}\right\}.
\end{equation}
Hence, from now on, whenever we write $\widehat{F}_R(y)$ it is this adapted estimator we refer to.

One of the key ideas we will be working on involves sampling from the distribution $\widehat{F}_R$. We can do this via the inverse distribution function approach; i.e., for a random uniform variable $u$ from $(0,1)$ we take the sample as
\begin{equation}\label{sample}
\arg\inf_y\,\, \widehat{F}_R(y)=u.
\end{equation}

\vspace{0.1in}
\noindent
{\sc Example 1.} In the first toy example we take $n=100$ and take the data $Y_{1:n}$ as independent standard normal random variables. We plot $\hatt{F}_R(y)$ in Figure~\ref{fig:cumulative_distribution}(a) with $R=5$. We then sample 10000 variables from $\hatt{F}_R$ and these are represented as a histogram in Figure~\ref{fig:cumulative_distribution}(b). As we can see from these figures, both the estimated cumulative distribution and the histogram respectively yield good estimation of the true cumulative distribution and the density function of the standard normal random variable.
\begin{figure}[!t]
\includegraphics[width=14cm,height=8cm]{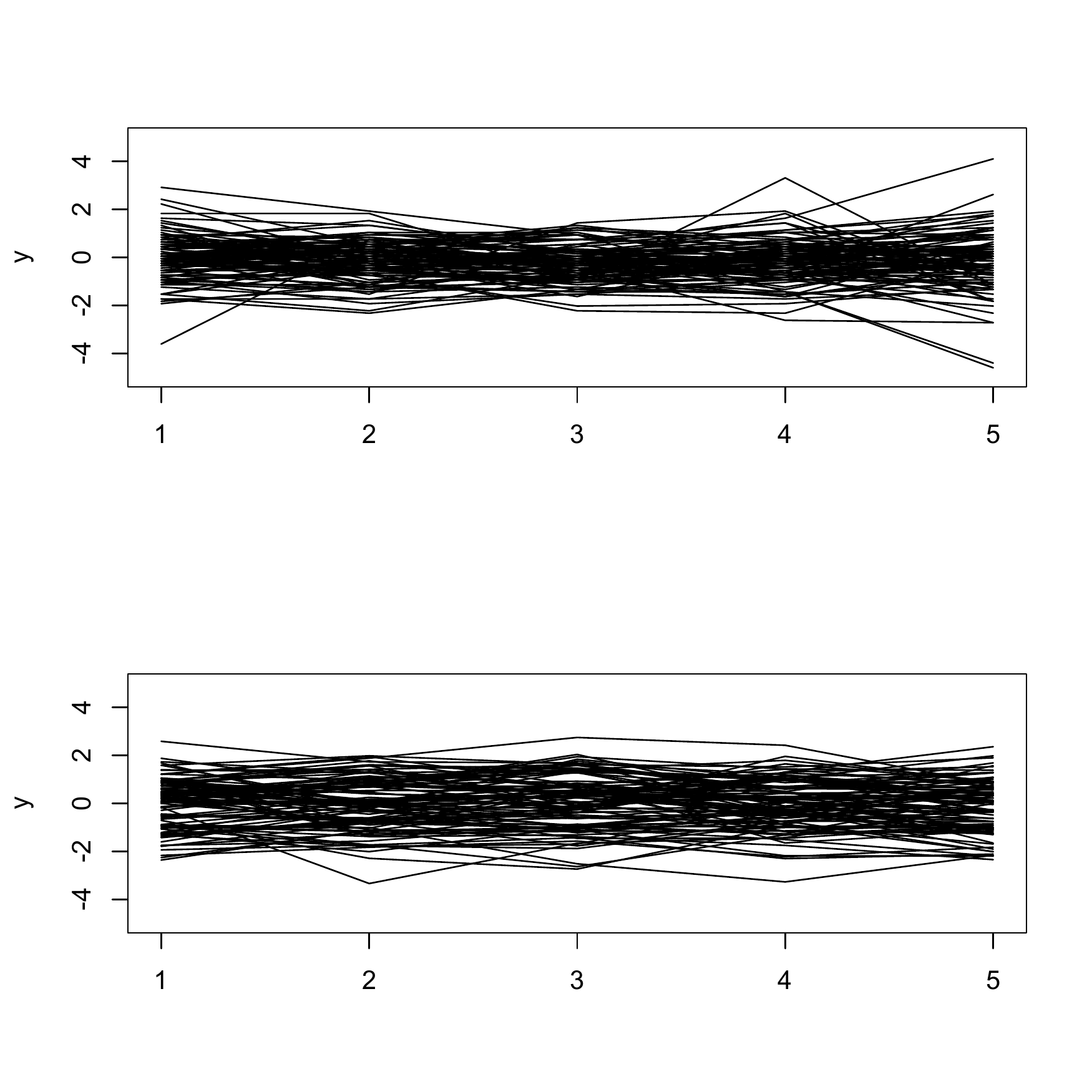}
\caption{Generated samples (top) from estimated conditional distribution and 100 data samples (bottom) from the multivariate normal distribution with a Markov structure in Example 3.}
\label{fig4}
\end{figure}

\noindent
{\sc Example 2.} In the next example we consider a conditional distribution function with two conditional variables.
We take $n=500$ independent observations from a three sequence of Markov variables;
$$y_1=\N(0,1),\quad [y_2\mid y_1]=\N(\rho y_1,1-\rho^2),\quad [y_3\mid y_2]=\N(\rho y_2,1-\rho^2)$$
and take $\rho=0.6$. We estimate the conditional distribution of $y_2$ given $y_1=y_3=0$ which in general is
$$f(y_2\mid y_1,y_3)=\N\left(\frac{c(y_1+y_3)}{1+c^2},\frac{1-c^2}{1+c^2}\right).$$
The estimated distribution is a straightforward extension of the one dimensional case, which is given by:
\begin{align*}
\hatt{F}_{R_1,R_2}(y_2\mid y_1,y_3)=
\half+\frac{1}{\pi}\frac{\sum_{i=1}^n \Si(R_1(y_2-y_{2i}))\,K_{R_2}(y_1-y_{1i})K_{R_2}(y_3-y_{3i})}
{\sum_{i=1}^n K_{R_2}(y_1-y_{1i})K_{R_2}(y_3-y_{3i})}.
\end{align*}
where we are now writing
$K_R(z) : =\sin(Rz)/z.$
We also allow for the $R$ to be different depending on its placement within the estimator; i.e., with the dependent or conditional variables.

The estimated distribution along with the true distribution are shown in Figure~\ref{fig3}. For this we took $R_1=10$ and $R_2=6$. As usual we implemented adaption (\ref{minmax}). For graphical representations when drawing the distribution using a grid we implement the isotonic regression technique to ensure the function is non--decreasing. However, for sampling, this is not necessary. 
\begin{figure}[!t]
\includegraphics[width=14cm,height=9cm]{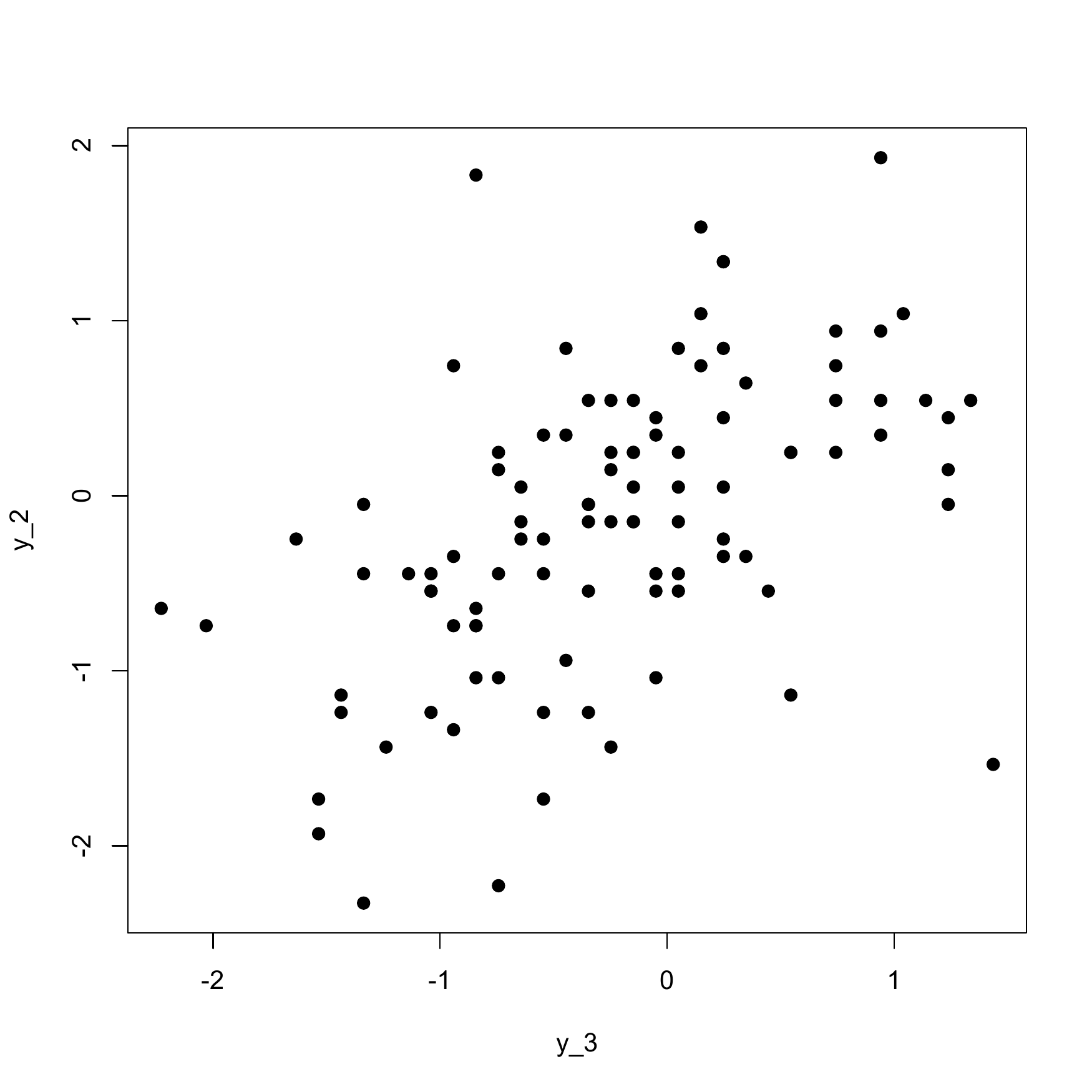}
\caption{Plot of generated samples $(y_2^*, y_3^*)$ in Example 3.}
\label{fig5}
\end{figure}
\vspace{0.2in}
\noindent
For independent observations $y$ from some $d$--dimensional distribution function $F$ we can
provide a generative model by estimating a sequence of marginal and conditional distribution functions.
That is, suppose $d=5$ and we  extend the set--up illustrated in Example 2. 

\vspace{0.2in}
\noindent
{\sc Example 3.}
We take data with a sample size of $n=1000$ from the $d=5$ multivariate normal distribution with a Markov structure; so we take 
 $y_1=\N(0,1)$ and for $j=2,\ldots,d$ we have
$[y_j\mid y_{j-1}]=\N\left(\rho y_{j-1},1-\rho^2\right).$
We use $\rho=0.6$. 

For the generative estimator we assume we do not know the data structure. We do this by obtaining marginal and conditional distributions; specifically we estimate
$$\hatt{F}(y_1)=\half+\frac{1}{n\pi}\sum_{i=1}^n \Si(R_1(y_1-y_{1i}))$$
and for $j=2:d$, 
$$\hatt{F}(y_j\mid y_{1:j-1})=\half+\frac{1}{\pi}\frac{\sum_{i=1}^n 
\Si(R_1(y_j-y_{ji}))\,\prod_{l=1}^{j-1} K_{R_2}(y_l-y_{li})}
{\sum_{i=1}^n \prod_{l=1}^{j-1} K_{R_2}(y_l-y_{li})}.$$ 
To generate samples we use the idea from equation~(\ref{sample}) sequentially, starting with $\hatt{F}(y_1)$ to get $y_{1}^*$ and subsequently $y_{2:d}^*$ using 
the $\hatt{F}(y_j\mid y_{1:j-1}^*)$.

Figure~\ref{fig4} shows 100 generated $y_{1:d}^*$ and also 100 of the 1000 $y_{1:d}$ represented as a time series plot. When viewed as samples from a joint distribution, each $y_j$ is marginally standard normal and has correlation 0.6 with $y_{j-1}$. The average of the means of the generated samples is -0.15 and the average of the 
variances is 0.96.
A plot of the generated samples $(y_2^*,y_3^*)$ is presented in Figure~\ref{fig5} and the measured correlation between the pairs of samples is 0.50.
\begin{figure}[t!]
\includegraphics[width=14cm,height=9cm]{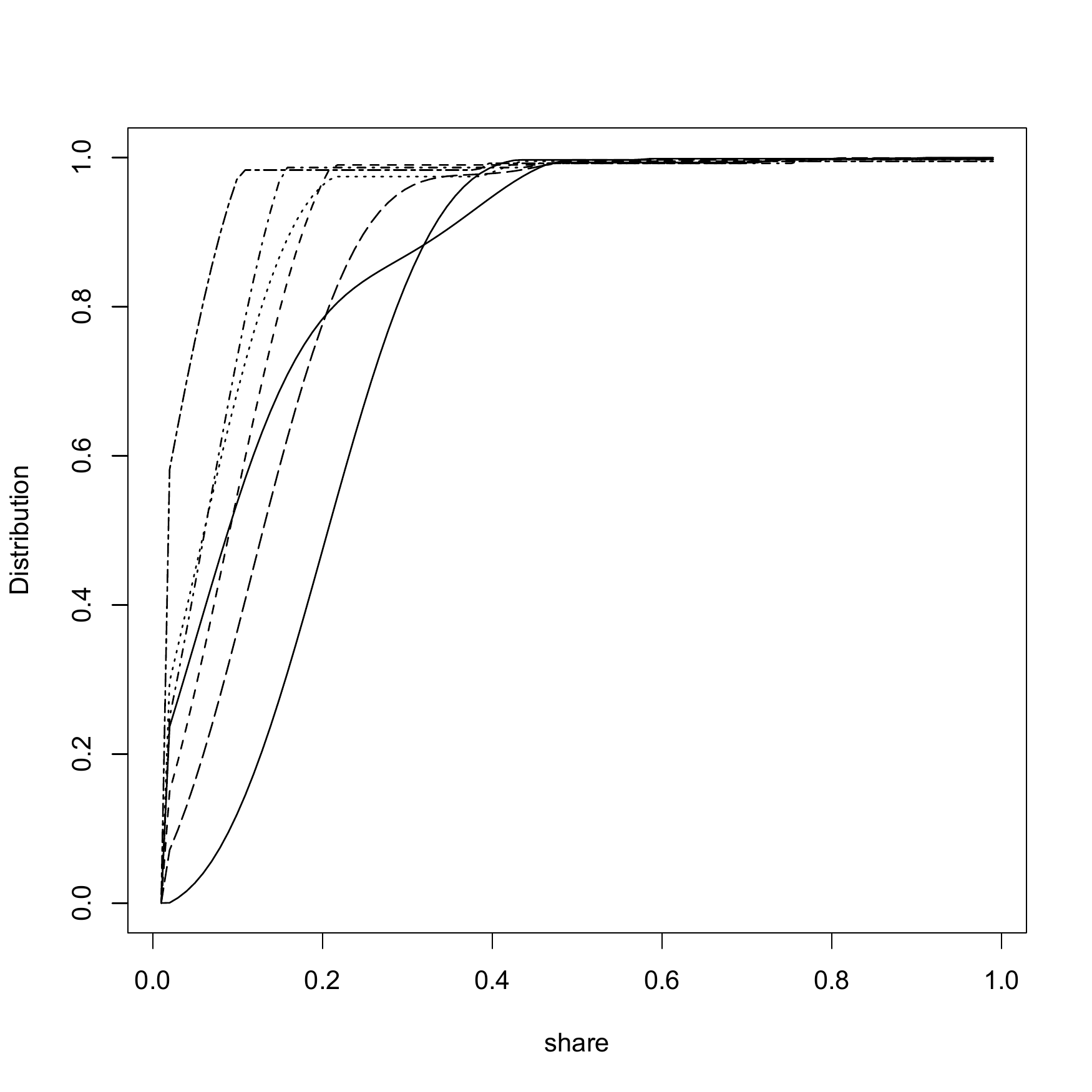}
\caption{Sequence of conditional distributions in the Engel95 dataset in Example 4.} 
%{\color{blue} Can we use different colors for each curve?}}
\label{fig13}
\end{figure}

\vspace{0.1in}
\noindent
{\sc Example 4.}
A further example involves a dataset from the R package \textbf{np} and is described in \cite{Hayfield08}. The dataset is ``Engel95'' and consists of household data from 1655 married families with 10 observations per family including expenditure on food, catering, fuel and other such commodities. One of the variables is the number of children in the household which is discrete and we leave that variable out of the dataset. We remain with 9 variables and condition the 7  expenditure shares of  food, catering, alcohol, fuel, motor, fares, and leisure on the log of total expenditure and the log of total earnings. We number the variables in the same order as we have just written them here.  

We do the inference via sequential conditioning, starting with the conditional for $F(y_1\mid y_8,y_9)$ and then the conditionals $F(y_l\mid y_{1:l-1},y_8,y_9)$ for $l=2,\ldots,7$. We estimate the conditional distributions at the means of each of the variables and the conditionals are presented in Figure~\ref{fig13}.
The jumps at the start of some of the distributions are due to a number of households recording zero share in some of the expenditures. In the analysis, we took $R=20$.
\section{Quantile regression and Markov data}
\label{Sec:Quantile_Markov}
In this section, we discuss an application of the Fourier integral theorem to quantile regression and Markov data.
\subsection{Quantile regression}
Quantile regression has become an important area of statistical analysis since the prioneering work of \cite{Koenker1978}. The most popular approaches to nonparametric quantile regression use the ``pinball'' loss function
$$l_u(\xi)=\left\{\begin{array}{ll}
u\xi  & \xi\geq 0 \\
(u-1)\xi & \xi<0,
\end{array}
\right.
$$   
with $0<u<1$ representing the quantile of interest. The idea is that the quantile regression function
$Q(u\mid x)$ is the solution to the minimization problem;
$$\min_f \sum_{i=1}^n l_u(y_i-f(x_i))+\lambda ||f||$$
where the observed data are $(x_i,y_i)_{i = 1}^{n}$ and $||\cdot||$ is some chosen penalty function while $f$ is modelled in traditional ways using for example splines or polynomials or kernels.  See, for example, \cite{Yu1998}, \cite{Liu2011}, \cite{Takeuchi2005}, and \cite{Koenker2005}. 
\begin{figure}[!t]
\includegraphics[width=14cm,height=9cm]{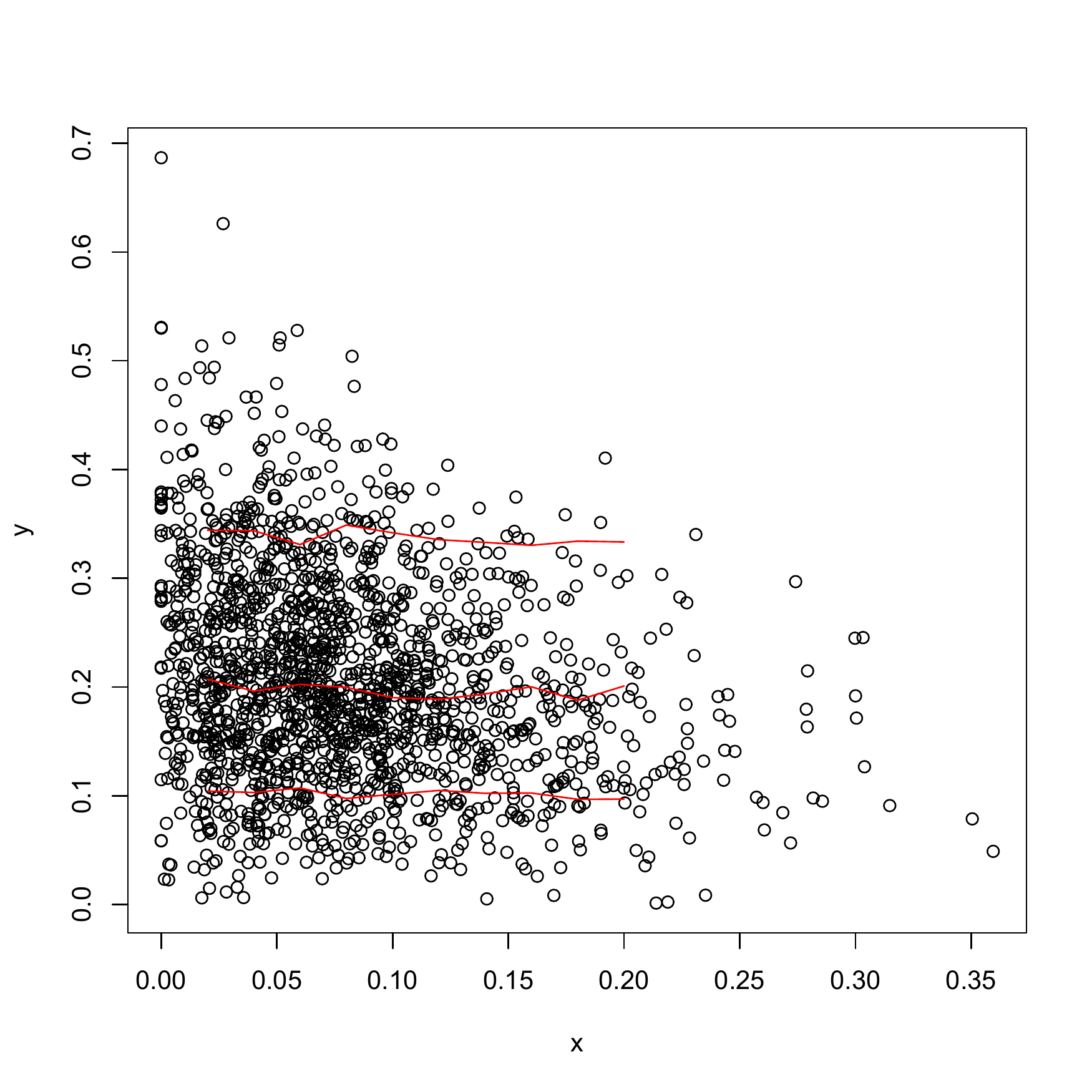}
\caption{Data plot with quartile functions from Engel95 dataset.}
\label{fig14}
\end{figure}

Another traditional approach is to get the inverse of a kernel estimated distribution function, e.g., \cite{Daouia2013}; noting that there appears no direct kernel estimation of a quantile. However, here we show how the Fourier integral theorem can be used to obtain a quantile regression function using the $K_R$ kernel. We are looking  to find the function $Q(u\mid x)$, the quantile function corresponding to the distribution $F(y\mid x)$.
Now, fixing $R$, we have
$$Q_R(u\mid x)=\frac{1}{\pi}\int \frac{\sin(R(u-v))}{u-v}\,Q(v\mid x)\,d v.$$
Transforming $v=F(y\mid x)$ we get
$$Q_R(u\mid x)=\frac{1}{\pi}\int \frac{\sin(R(u-F(y\mid x)))}{u-F(y\mid x)}\,y\,f(y\mid x) d y.$$
Consequently, the Monte Carlo estimator of $Q_R(\cdot\mid x)$ is given by
$$\hatt{Q}_R(u\mid x)=\frac{1}{n\pi}\sum_{i=1}^n \frac{\sin(R(u-\hatt{F}(y_i^*\mid x)))}{u-\hatt{F}(y_i^*\mid x)}\,y_i^*,$$
where the $(y_i^*)$ are taken from $\hatt{F}(\cdot\mid x)$, which itself is estimated as
$$\hatt{F}(y\mid x)=\half+\frac{1}{\pi}\frac{\sum_{j=1}^n \Si(R(y-y_j))\,K_R(x-x_j)}
{\sum_{j=1}^n K_R(x-x_j)}.$$

We took data from the Engel95 data set; the first and second columns, regressing the first column on the second column. With $n=1655$ we took $R=10$ and took 5000 samples from each $\hatt{F}(\cdot\mid x)$ with the $x$ values being $(0.02,0.04,\ldots,0.2)$. We then computed the quartiles; the data with the quartile functions are presented in Figure~\ref{fig14}.
\begin{figure}[t]
\centering
\begin{subfigure}[t]{0.48\textwidth}
\includegraphics[width=1\textwidth]{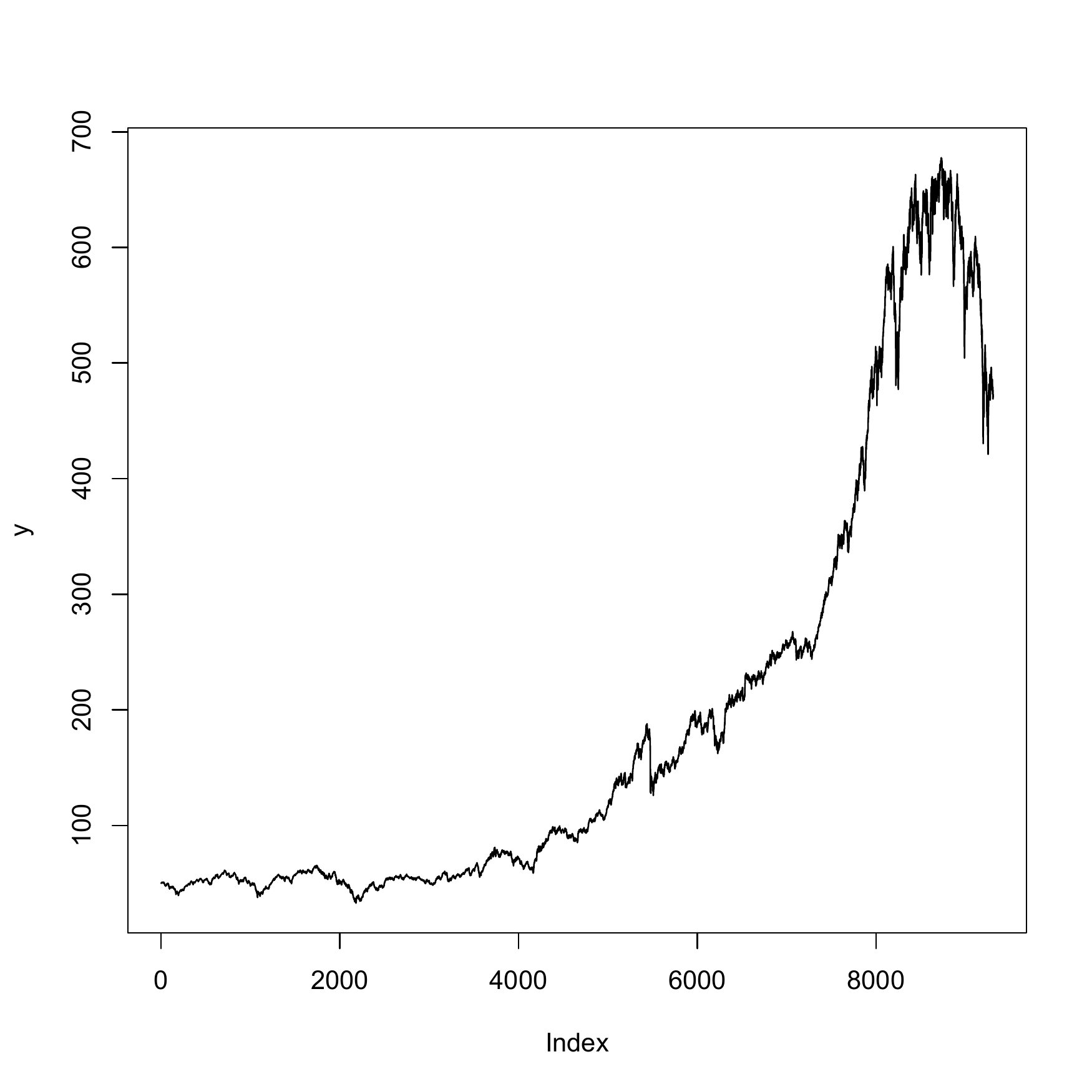}
\caption{}
\end{subfigure}
\begin{subfigure}[t]{0.48\textwidth}
\includegraphics[width=1\textwidth]{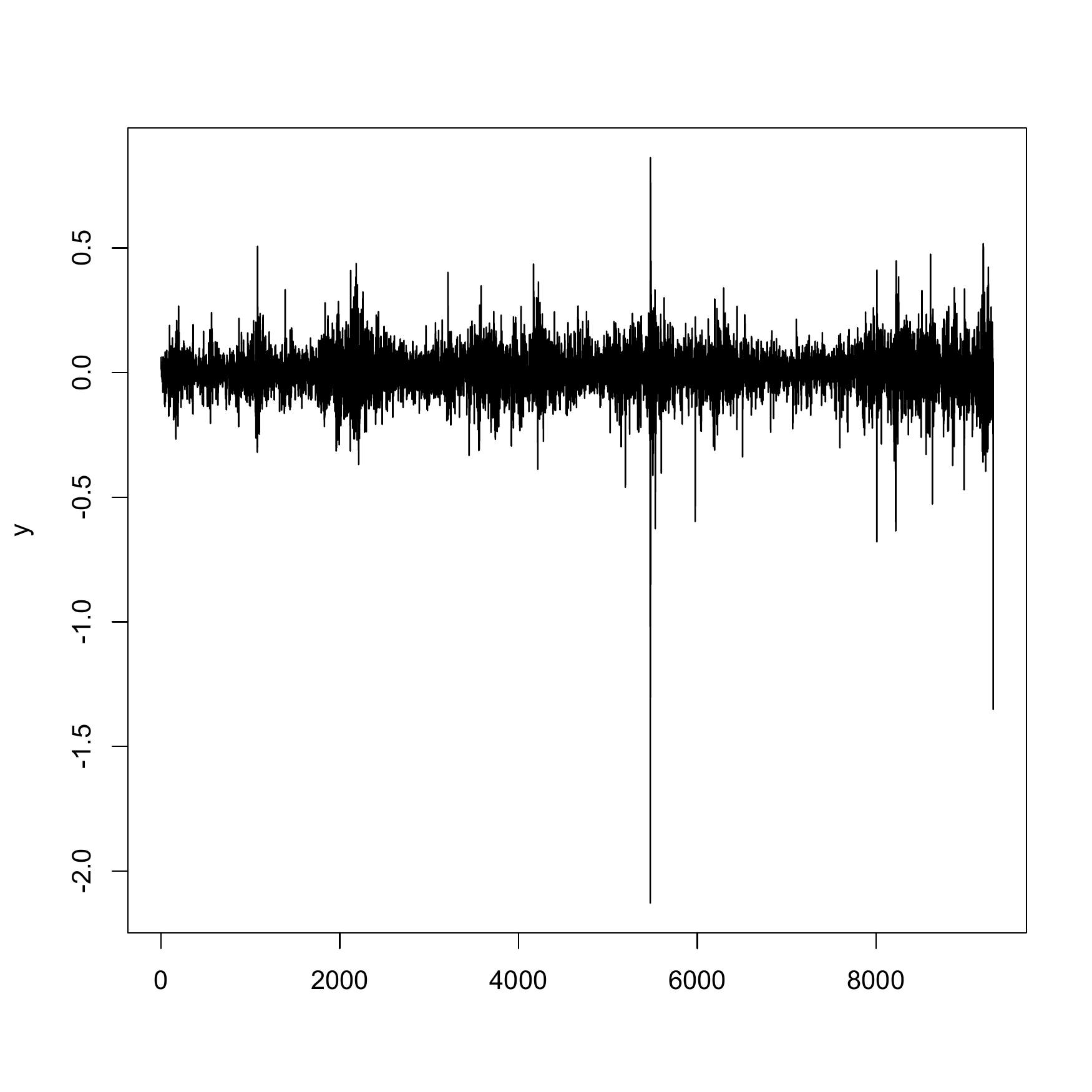}
\caption{}
\end{subfigure}
\caption{(a) Raw data of 9311 daily records of NYSE Composite Index; (b) Transformed NYSE Composite Index data.
}
\label{fig:markov_data}
\end{figure}
\subsection{Markov data}

The data for this section is to be found in the R package \emph{fBasics} and consists of $n=9311$ data points of daily records of the NYSE Composite Index. A plot of the raw data is given in Figure~\ref{fig:markov_data}(a). 
We analyze the transformed data
$y_i=10\log(yraw_{i+1}/yraw_i)$, where $(yraw_i)$ are the raw data. This gives us a sample size of $n=9310$. 

%\begin{figure}[t]
%\begin{center}
%\includegraphics[width=14cm,height=6cm]{Fig6}
%\caption{Raw data of 9311 daily records of NYSE Composite Index.}
%\label{fig6}
%\end{center}
%\end{figure}

%\begin{figure}[t]
%\begin{center}
%\includegraphics[width=14cm,height=6cm]{Fig7}
%\caption{Transformed NYSE Composite Index data.}
%\label{fig7}
%\end{center}
%\end{figure}
The aim is to assume a missing observation, we select the $m=1000$-th observation, and to generate samples for it. We do this in a full nonparametric setting where we make no assumption on the transition mechanism and most importantly do not assign a dependence structure from one observation to the next. The only assumption is the temporal homogeneity of the transformed data $(y_i)$. A plot of the transformed data supporting this assumption is provided in Figure \ref{fig:markov_data}(b).  

\begin{figure}[t]
\includegraphics[width=14cm,height=6cm]{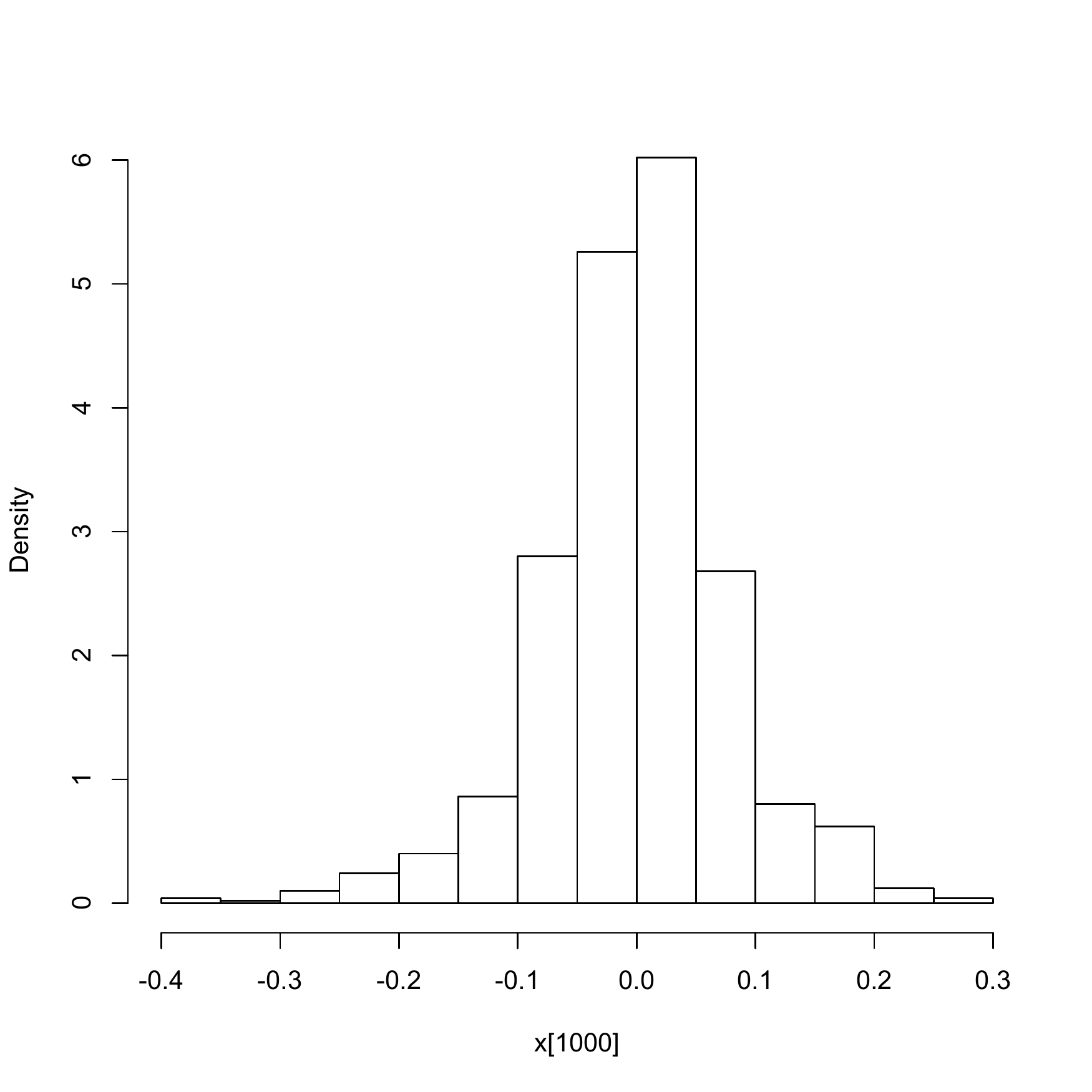}
\caption{Histogram of generated samples for $x[1000]$.}
\label{fig8}
\end{figure}

To impute samples for the missing observation, we rely on the Markov property and hence we need to estimate
$F(y_m\mid y_{m-1},y_{m+1}).$
This is provided by
\begin{align*}
\hatt{F}(y_m\mid y_{m-1},y_{m+1}) & = \half \\
& \hspace{- 3 em} + \frac{1}{\pi}\frac{\sum_{i\ne m-1,m,m+1}
\Si(R_1(y_m-y_i))\,K_{R_2}(y_{m-1}-y_{i-1})\,K_{R_2}(y_{m+1}-y_{i+1})}
{\sum_{i\ne m-1,m,m+1} K_{R_2}(y_{m-1}-y_{i-1})\,K_{R_2}(y_{m+1}-y_{i+1})}.
\end{align*}
For the simulation we took $R_1=R_2=50$  and took 1000 generated samples from the distribution estimator. A histogram of the samples is presented in Figure~\ref{fig8}. The true value of $x[1000]$ is 0.12 and the values either side are $x[999]=0.080$ and $x[1001]=-0.070$. So the value 0.12 is not lying in between the two where a lot of the mass is, quite rightly, from the samples. Nevertheless, the coverage of 0.12 is good.

\section{Mixing distribution}
\label{sec:mixing}
In this section we consider a nonparametric regression model where the density for observations $y$ given $x$
is given by
$p(y\mid x)=\int K(y-\theta)\,d G(\theta\mid x),$
for some kernel $K$, which we assume to be normal distribution with variance $h^2$.
The aim is to estimate $G(\theta\mid x)$ for a new predictor $x$. We will write out the procedure assuming $x$ is one dimensional but this is easily extendable to higher dimension. 
To get to the desired estimator, we first get the conditional distribution $P(y\mid x)$ using
$$\hatt{P}_{\radius_{1}, \radius_{2}}(y\mid x)=\half +\frac{1}{\pi}\frac{\sum_{i=1}^n \Si(R_1(y-y_i))\,K_{R_2}(x-x_i)}
{\sum_{i=1}^n K_{R_2}(x-x_i)}.$$
We then get the estimator for $G(\theta\mid x)$ using $\hatt{P}_{\radius_{1}, \radius_{2}}(y\mid x)$, which is given by:
$$\widehat{G}_{\radius, \radius_{1}, \radius_{2}}(\theta\mid x)=\half+\frac{1}{\pi}\int_0^R\,\int_{-\infty}^\infty e^{\half s^2h^2}\,\frac{\sin(s(\theta-y))}{s}\,d \hatt{P}_{\radius_{1}, \radius_{2}}(y\mid x)\,d s.$$

In practice, for a new predictor $x$, we sample the data $(y_{i}^{*})_{i = 1}^{N}$ from $\hatt{P}_{\radius_{1}, \radius_{2}}(y\mid x)$. The best way to use the samples $(y_i^*)$ and to complete the $s$ integral is that we sample alongside the $y_i^*$ an independent $s_i$ taken uniformly from the interval $(0,R)$. Hence, we get as our final estimator,
\begin{equation}\label{mixing}
\hatt{G}(\theta\mid x)=\half+\frac{R}{N\pi}\sum_{i=1}^N e^{\half s_i^2 h^2}\,\frac{\sin(s_i(\theta-y_i^*))}{s_i}.
\end{equation}
For this estimator we are at liberty to select our own sample size $N$. 

\subsection{Simulated data}

\begin{figure}[t]
\centering
\begin{subfigure}[t]{0.48\textwidth}
\includegraphics[width=1\textwidth]{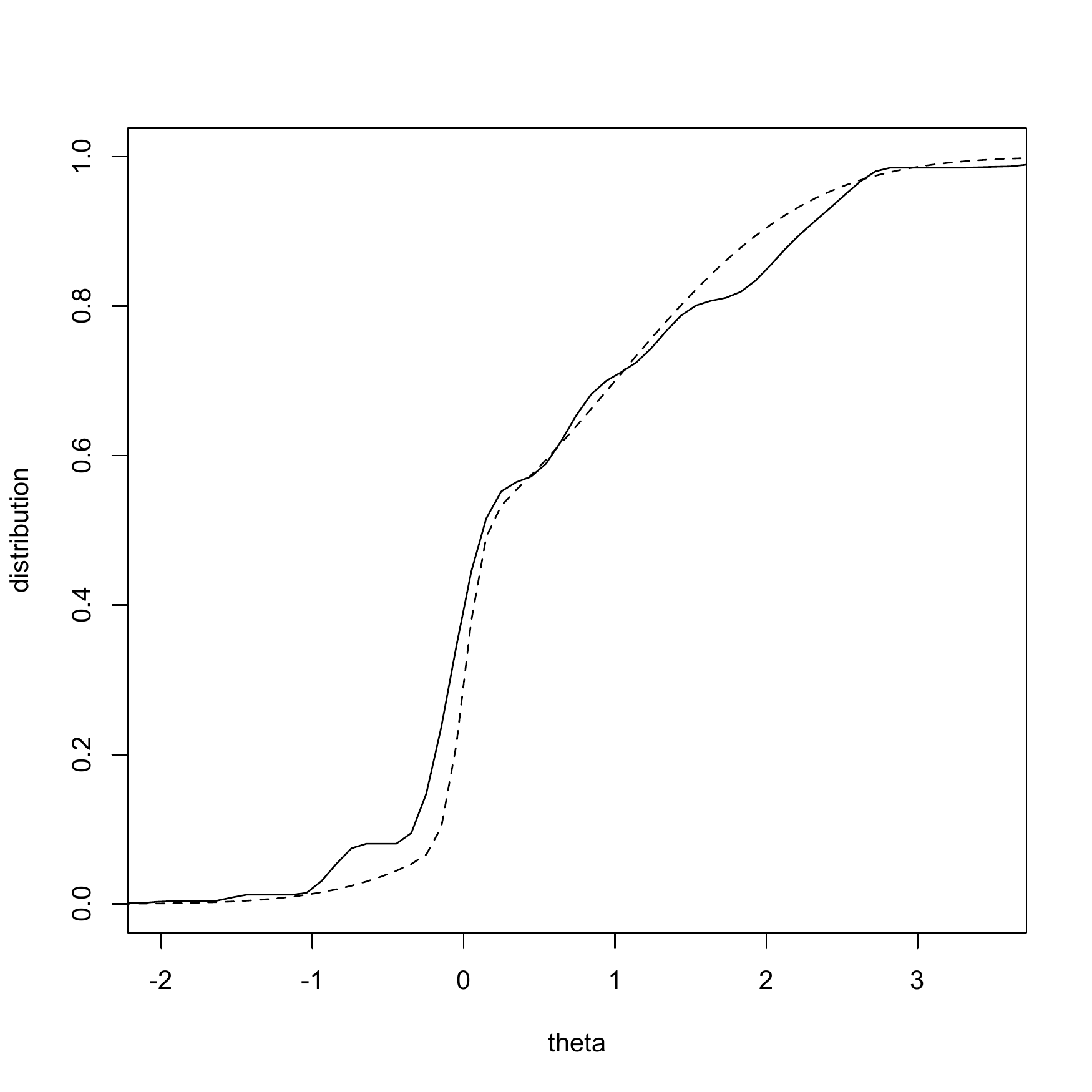}
\caption{}
\end{subfigure}
\begin{subfigure}[t]{0.48\textwidth}
\includegraphics[width=1\textwidth]{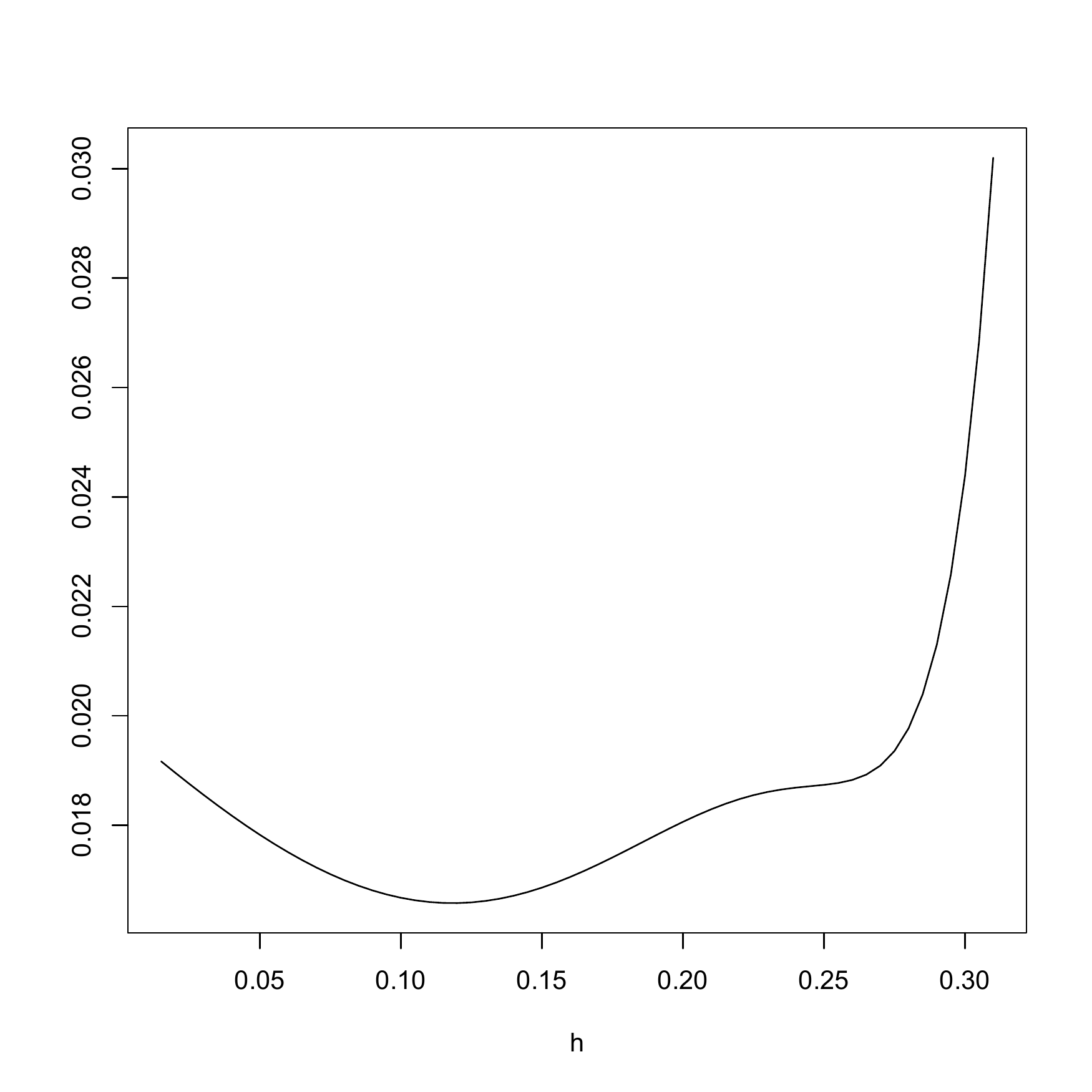}
\caption{}
\end{subfigure}
\caption{(a) Estimate of $G(\theta\mid x=1)$ (bold line) alongside true distribution (dashed line); (b) Plot of $\alpha(h)$ vs $h$.
}
\label{fig:mixing_distribution}
\end{figure}

We first illustrate with some simulated data; with $n=1000$ we take observations from the mixture model with
$$g(\theta\mid x)=0.4\,\N(\theta\mid 0, 0.1^2x^2)+0.6\,\N(\theta\mid x,1).$$
The $(x_i)$ are sampled as standard normal and for the normal kernel we take $h=0.1$.
We then estimate $G(\theta\mid x)$ with $x=1$ and take $R=R_1=R_2=10$ throughout. A plot of the estimated distribution alongside  the true one is given in Figure~\ref{fig:mixing_distribution}(a).

%\begin{figure}[t]
%\begin{center}
%\includegraphics[width=14cm,height=6cm]{Fig9}
%\caption{Estimate of $G(\theta\mid x=1)$ (bold line) alongside true distribution (dashed line).}
%\label{fig9}
%\end{center}
%\end{figure}
%\begin{figure}[t]
%\begin{center}
%\includegraphics[width=14cm,height=6cm]{Fig10}
%\caption{Plot of $\alpha(h)$ vs $h$.}
%\label{fig10}
%\end{center}
%\end{figure}
In general, the value for $h$ will not be known. However, it is straightforward to estimate and we outline the procedure here. The key point is that
$$P(y\mid x)=\int G_h(y-h\,z\mid x)\,\phi(z)\,d z.$$
We estimate $P(y\mid x)$ without any reference to $h$. Hence, we can estimate $G_h(\cdot\mid x)$ for a range of $h$ and select the $h$ which minimizes
$$\alpha(h)=\left|P(y\mid x)-\int G_h(y-h\,z\mid x)\,\phi(z)\,d z\right|.$$
There is no need to do this for all $(x,y)$; in practice we can select specific values for $x$ and $y$ such as the respective sample means. The evaluation of the integral of $G$ with respect to the standard normal density can be done using Monte Carlo methods. An illustration involved us taking a linear model for $G(\cdot\mid x)$; i.e. 
$G(\theta\mid x)=\Phi(\theta-x)$, and the plot of the $\alpha(h)$ verses $h$ is given in Figure~\ref{fig:mixing_distribution}(b). The minimum is at $h=0.12$ and the true value is $0.1$. 

\subsection{Real data}

The aim in this subsection is to highlight the performance of the mixing distribution estimator while modeling covariates as an alternative to standard nonparametric curve estimation. A common model for nonparametric regression is of the form
$$y_i=m(x_i)+h\epsilon_i$$
where $\epsilon_i$ are assumed to have zero mean and unit variance. In many cases, such as a dataset we will consider, the variance is nonhomogeneous making estimation of $m(\cdot)$ problematic.

An alternative way to model such data is via the mixture model;
$$y_i=\theta_i+h\epsilon_i$$
where we assume the $(\epsilon_i)$ are standard normal and we then model the 
$\theta_i$ to be from $G(\cdot\mid x_i)$, which we estimate using equation~(\ref{mixing}). For estimating the curve, we can 
numerically find the median of $\hatt{G}(\cdot\mid x)$ and use this as an estimator.

\begin{figure}[t]
\includegraphics[width=14cm,height=6cm]{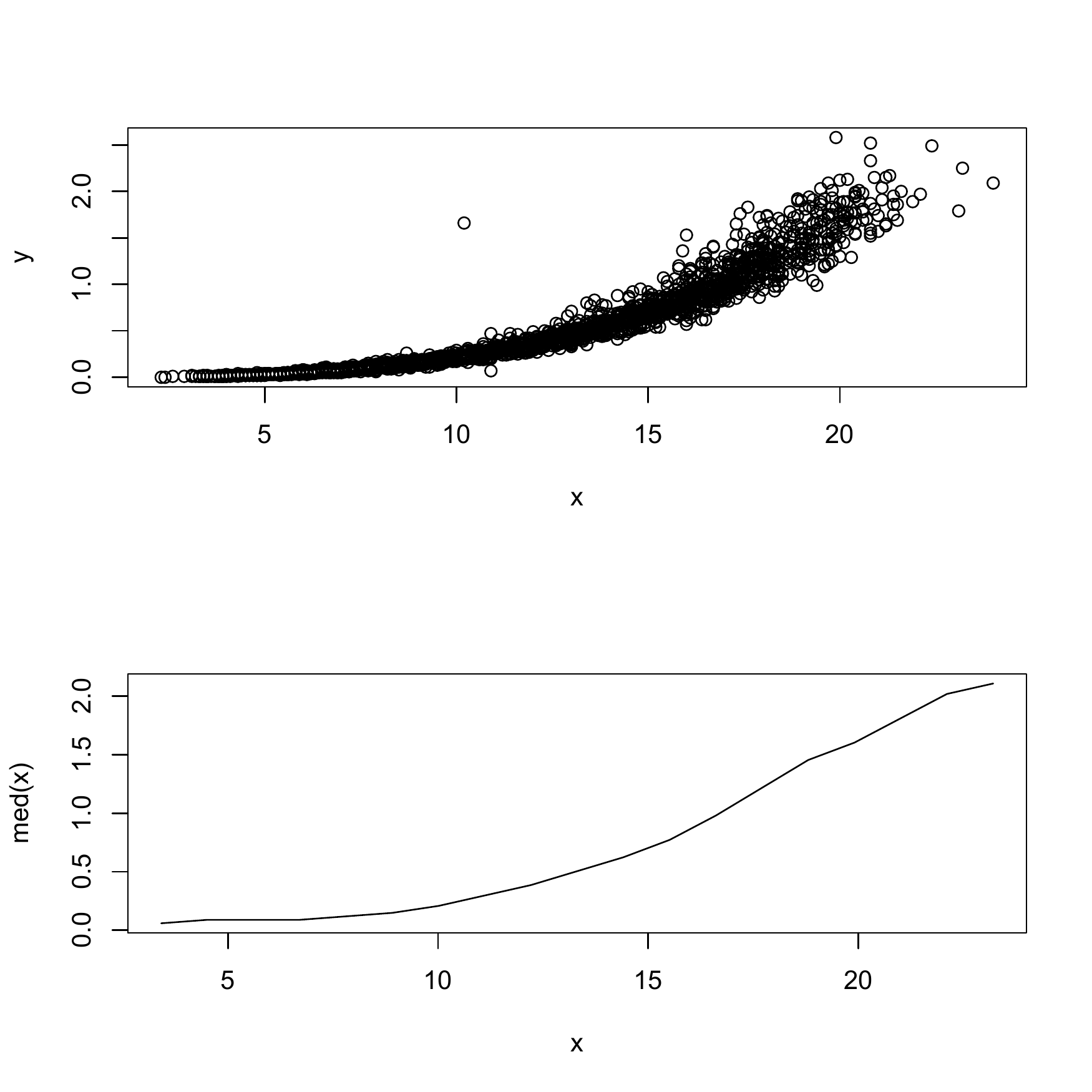}
\caption{Barnacle data (top) and median estimator of curve from mixing distribution (bottom).}
\label{fig12}
\end{figure}

To demonstrate this we use a dataset from the R package \textbf{npregfast} and a description of the ``barnacle'' data appears in \cite{Sestelo17}. Briefly here, we model $y$ as the dry weight of barnacles, taken from the Atlantic, and $x$ is the rostrocarinal length. The sample size is $n=2000$ and we model the data using $R=10$ and setting $h$ to be 0.05. A plot of the data and the corresponding median curve is presented in Figure~\ref{fig12}.

\section{Discussion}
\label{sec:discussion}
In this paper we have used the Fourier integral theorem and associated Monte Carlo estimators to provide new nonparametric estimators for conditional distribution functions and related functions, such as quantile functions. Indeed the quantile regression function used in Section 3 is novel.  The integral theorem is potentially a powerful tool which has yet to be utilized in statistical analysis; relevant as it provides natural Monte Carlo estimators of functions.

%assume that $X_{1}, \ldots, X_{n} \in \mathcal{X} \subseteq \mathbb{R}$ are i.i.d. samples from the density function $p_{0}$. Our goal is to estimate the cumulative distribution function of $p_{0}$. We propose the following estimator:
%\begin{align}
%    \cumest_{n, \radius}(x) = \frac{1}{2} + \frac{1}{\pi n} \sum_{i = 1}^{n}
%\int_{0}^{\radius} \frac{\sin(s(x - X_{i}))}{s} ds, \label{eq:Fourier_cdf_estimation}
%\end{align}
%for any radius $\radius > 0$. We refer the above estimator to as \emph{Fourier cumulative estimator}. The above estimation is motivated by the following representation of cumulative distribution function $F$:
%\begin{align}
%    F(x) = \frac{1}{2} + \frac{1}{\pi} \int_{0}^{\infty} \int_{-\infty}^{\infty} \frac{\sin(s(x - t))}{s} p_{0}(t) dt ds. \label{eq:cdf_representation}
%\end{align}
%It is clear from the definition of the Fourier cumulative estimator that $\lim_{x \to -\infty} \cumest_{n, \radius}(x) = 0$ and $\lim_{x \to \infty} \cumest_{n, \radius}(x) = 1$. Furthermore, the function $\cumest$ is also right-continuous. The issue with the Fourier cumulative estimator is that it is not non-decreasing and can also be negative or above 1. For the non-decreasing issue, we can utilize the idea of isotonic regression. Then, we can take maximum and minimum of the new values respectively with 1 and 0 to resolve the latter issue. 
%%%%%%%%%%%%%%%%%%%%%%%%%%%%%%%%%%%%%%%%%%%%%%%%%%%%%%%%%%%%%%%%%%%%%%%%%%
\section{Appendix}
\label{sec:Appendix}
In this appendix, we give the proof of Theorem~\ref{theorem:approximation_Fourier}. To ease the presentation, the values of universal constants (e.g., $C$, $C_{1}$, $C_{2}$, $\bar{C}$ etc.) can change from line-to-line. For any $x \in \mathbb{R}^{d}$, we denote $x = (x_{1}, \ldots, x_{d})$. 
\subsection{Proof of Theorem~\ref{theorem:approximation_Fourier}}
We first prove the result of Theorem~\ref{theorem:approximation_Fourier} when $d = 1$. In particular, we would like to show that when the function $m \in \mathcal{T}^{K}(\mathbb{R})$, there exists a universal constant $C > 0$ such that we have
\begin{align*}
	\abss{m_{R}(y) - m(y)} \leq \frac{C}{R^{K}}.
\end{align*}
In fact, from the definition of $m_{R}(y)$ in equation~\eqref{eq:Fourier_approx}, we have
\begin{align*}
    \abss{m_{R}(y) - m(y)} = \abss{\frac{1}{\pi} \int_{\mathbb{R}} \frac{\sin(R(y-x))}{(y - x)} \parenth{m(x) - m(y)} dx}.
\end{align*}
For simplicity of the presentation, for any $y \in \mathbb{R}$ we write $g(x) = (m(x) - m(y))/(x - y)$ for all $x \in \mathbb{R}$. Then, we can rewrite the above equality as 
\begin{align*}
    \abss{m_{R}(y) - m(y)} & = \abss{\frac{1}{\pi} \int_{\mathbb{R}} \sin(R(y-x)) g(x) dx} \\
    & = \abss{\frac{1}{\pi} \sum_{k = -\infty}^{\infty} \int_{y + \frac{2\pi k}{R}}^{y + \frac{2\pi(k + 1)}{R}} \sin(R(y-x)) g(x) dx}.
\end{align*}
Invoking the change of variables $x = y + \frac{t + 2 \pi k}{R}$, the above equation becomes
\begin{align}
    \abss{m_{R}(y) - m(y)} & = \abss{\frac{1}{\pi R} \sum_{k = -\infty}^{\infty} \int_{[0, 2\pi)} \sin(t) \cdot g \parenth{y + \frac{t + 2 \pi k}{R}}dt}. \label{eq:key_equation}
\end{align}
Since $m \in \mathcal{T}^{K}(\mathbb{R})$, the function $g$ is differentiable up to the $K$-th order. Therefore, using a Taylor expansion up to the $K$-th order, leads to 
\begin{align*}
    g \parenth{y + \frac{t + 2 \pi k}{R}} & = \sum_{\alpha \leq K - 1} \frac{1}{R^{\alpha}} \frac{t^{\alpha}}{\alpha!} g^{(\alpha)} \parenth{y + \frac{2 \pi k}{R}} \\
    & \hspace{4 em} + \frac{t^{K}}{R^{K} (K - 1)!} \int_{0}^{1} (1 - \xi)^{K - 1} g^{(K)} \parenth{y + \frac{2 \pi k}{R} + \frac{\xi t}{R}} d \xi.
\end{align*} 
Plugging the above Taylor expansion into equation~\eqref{eq:key_equation}, we have
\begin{align}
    \abss{m_{R}(y) - m(y)} = \abss{\frac{1}{\pi R} \sum_{\ell = 0}^{K} A_{\ell}} \leq \frac{1}{\pi R} \sum_{\ell = 0}^{K} \abss{A_{\ell}}, \label{eq:key_inequality_Fourier}
\end{align}
where, for $\ell \in \{0,1, \ldots, K - 1\}$, we define
\begin{align*}
    A_{\ell} & =  \frac{1}{R^{\ell}} \int_{[0, 2\pi)} \parenth{\frac{t^{\ell} \sin(t)}{\ell !}} dt \parenth{ \sum_{k = -\infty}^{\infty} g^{(\ell)} \parenth{y + \frac{2 \pi k}{R}}}, \ \text{and} \\
    A_{K} & = \sum_{k = -\infty}^{\infty} \int_{[0, 2\pi)} \sin(t) \parenth{ \frac{t^{K}}{R^{K}(K - 1)!} \int_{0}^{1} (1 - \xi)^{K - 1} g^{(K)} \parenth{y + \frac{2 \pi k}{R} + \frac{\xi t}{R}} d \xi} dt.
\end{align*}
We now find a bound for $\abss{A_{\ell}}$ for $\ell \in \{0,1,\ldots,K - 1\}$; we will demonstrate that
\begin{align}
    \abss{\sum_{k = -\infty}^{\infty} g^{(\ell)} \parenth{y + \frac{2 \pi k}{R}}} \leq \frac{C}{R^{K - (\ell + 1)}}, \quad \text{for all} \ \ell \in \{0,1,\ldots,K - 1\} \label{eq:inequality_first}
\end{align}
where $C$ is some universal constant. To obtain these bounds, we will use an inductive argument on $\ell$. We first start with $\ell = K - 1$. In fact, we have
\begin{align*}
    & \hspace{-5 em} \biggr|\sum_{k = -\infty}^{\infty} g^{(K - 1)} \parenth{y + \frac{2 \pi k}{R}} \frac{(2\pi)}{R} - \int_{\mathbb{R}} g^{(K - 1)}(x) dx \biggr| \\
    & \hspace{3 em} = \biggr|\sum_{k = -\infty}^{\infty} \int_{y + \frac{2\pi k}{R}}^{y + \frac{2\pi(k + 1)}{R}}  \biggr(g^{(K - 1)}(x) - g^{(K - 1)} \parenth{y + \frac{2 \pi k}{R}}\biggr)  dx \biggr| \\
    & \hspace{3 em} \leq \sum_{k = -\infty}^{\infty} \int_{y + \frac{2\pi k}{R}}^{y + \frac{2\pi(k + 1)}{R}} \abss{g^{(K - 1)}(x) - g^{(K - 1)} \parenth{y + \frac{2 \pi k}{R}}} dx.
\end{align*}
An application of Taylor expansion leads to
\begin{align*}
    g^{(K - 1)}(x) & = g^{(K - 1)} \parenth{y + \frac{2 \pi k}{R}} +  \parenth{x - y - \frac{2\pi k}{R}} \int_{0}^{1} g^{(K)} \parenth{(1 - \xi) \parenth{y + \frac{2 \pi k}{R}} + \xi x } d \xi.
\end{align*}
Now for any $\xi \in [0, 1]$ and $x \in \brackets{y + \frac{2\pi k}{R}, y + \frac{2 \pi (k + 1)}{R}}$, we have
\begin{align*}
    \abss{g^{(K)} \parenth{(1 - \xi) \parenth{y + \frac{2 \pi k}{R}} + \xi x }} \leq  \sup_{t \in \brackets{y + \frac{2\pi k}{R}, y + \frac{2 \pi (k + 1)}{R}}} \abss{g^{(K)}(t)}.
\end{align*}
Collecting the above results, we find that 
\begin{align*}
    & \hspace{- 3 em} \sum_{k = -\infty}^{\infty} \int_{y + \frac{2\pi k}{R}}^{y + \frac{2\pi(k + 1)}{R}} \abss{g^{(K - 1)}(x) - g^{(K - 1)} \parenth{y + \frac{2 \pi k}{R}}} dx \\
    & \leq \sum_{k = -\infty}^{\infty} \parenth{\int_{y + \frac{2\pi k}{R}}^{y + \frac{2\pi(k + 1)}{R}} \parenth{x - y - \frac{2\pi k}{R}} dx} \sup_{t \in \brackets{y + \frac{2\pi k}{R}, y + \frac{2 \pi (k + 1)}{R}}} \abss{g^{(K)}(t)} \\
    & = \frac{2 \pi^{2}}{R^{2}} \sum_{k = -\infty}^{\infty} \sup_{t \in \brackets{y + \frac{2\pi k}{R}, y + \frac{2 \pi (k + 1)}{R}}} \abss{g^{(K)}(t)}.
\end{align*}
Using a Riemann sum approximation theorem, we have
\begin{align*}
	\lim_{R \to \infty} \sum_{k = -\infty}^{\infty} \sup_{t \in \brackets{y + \frac{2\pi k}{R}, y + \frac{2 \pi (k + 1)}{R}}} \abss{g^{(K)}(t)} \frac{2\pi}{R} = \int_{\mathbb{R}} \abss{g^{(K)}(x)}dx < \infty,
\end{align*}
where the finite value of the integral is due to the assumption that $m \in \mathcal{T}^{K}(\mathbb{R})$. Furthermore, the above limit is uniform in terms of $y$ as $g^{(K)}$ is uniformly continuous. Collecting the above results, there exists a universal constant $C$ such that as long as $R \geq C$, the following inequality holds:
\begin{align*}
    \frac{2 \pi^{2}}{R^{2}} \sum_{k = -\infty}^{\infty} \sup_{t \in \brackets{y + \frac{2\pi k}{R}, y + \frac{2 \pi (k + 1)}{R}}} \abss{g^{(K)}(t)} \leq \frac{C_{1}}{R}
\end{align*}
where $C_{1}$ is some universal constant. Combining all of the previous results, we obtain 
\begin{align*}
    \biggr|\sum_{k = -\infty}^{\infty} g^{(K - 1)} \parenth{y + \frac{2 \pi k}{R}} \frac{(2\pi)}{R} - \int_{\mathbb{R}} g^{(K - 1)}(x) dx \biggr|  \leq \frac{C_{1}}{R}.
\end{align*}
Since $m \in \mathcal{T}^{K}(\mathbb{R})$, using integration by parts, we get
\begin{align*}
    \int_{\mathbb{R}} g^{(K - 1)}(x) dx= 0.
\end{align*}
Therefore, we obtain the conclusion of equation~\eqref{eq:inequality_first} when $\ell = K - 1$. 

Now assume that the conclusion of equation~\eqref{eq:inequality_first} holds for $1 \leq \ell \leq K - 1$. We will prove that the conclusion also holds for $\ell - 1$. With a similar argument to the setting $\ell = K - 1$, we  obtain 
\begin{align}
	    & \hspace{-5 em} \biggr|\sum_{k = -\infty}^{\infty} g^{(\ell)} \parenth{y + \frac{2 \pi k}{R}} \frac{(2\pi)}{R} - \int_{\mathbb{R}} g^{(\ell)}(x) dx \biggr| \nonumber \\
	    & \hspace{3 em} = \abss{\sum_{k = -\infty}^{\infty} \int_{y + \frac{2\pi k}{R}}^{y + \frac{2\pi(k + 1)}{R}} \parenth{g^{(\ell)}(x) - g^{(\ell)} \parenth{y + \frac{2 \pi k}{R}}} dx}. \label{eq:inductive_argument}
\end{align}
Using a Taylor expansion, we have
\begin{align*}
    g^{(\ell)}(x) & =  g^{(\ell)} \parenth{y + \frac{2 \pi k}{R}} + \sum_{\alpha \leq K - 1 - \ell} \frac{\parenth{x - y - \frac{2\pi k_{j}}{R}}^{\alpha}}{\alpha!} g^{(\ell + \alpha)} \parenth{y + \frac{2 \pi k}{R}} \\
    & + \frac{\parenth{x - y - \frac{2\pi k}{R}}^{K - \ell}}{(K - \ell - 1)!} \int_{0}^{1} (1 - \xi)^{K - \ell - 1} g^{(K)} \parenth{(1 - \xi) \parenth{y + \frac{2 \pi k}{R}} + \xi x } d \xi.
\end{align*}
Plugging the above Taylor expansion into equation~\eqref{eq:inductive_argument}, we find that
\begin{align*}
   \sum_{k = -\infty}^{\infty} \int_{y + \frac{2\pi k}{R}}^{y + \frac{2\pi(k + 1)}{R}} \abss{g^{(\ell)}(x) - g^{(\ell)} \parenth{y + \frac{2 \pi k}{R}}} dx \leq S_{1} + S_{2},
\end{align*}
where $S_{1}$ and $S_{2}$ are defined as follows:
\begin{align*}
    S_{1} & = \sum_{\alpha \leq K - 1 - \ell} \biggr|\sum_{k = -\infty}^{\infty} \parenth{\int_{y + \frac{2\pi k}{R}}^{y + \frac{2\pi(k + 1)}{R}} \frac{\parenth{x - y - \frac{2\pi k_{j}}{R}}^{\alpha}}{\alpha!} dx} g^{(\ell + \alpha)} \parenth{y + \frac{2 \pi k}{R}} \biggr| \\
    & = \sum_{\alpha \leq K - 1 - \ell}  \biggr|\sum_{k = -\infty}^{\infty} \frac{(2\pi)^{\alpha + 1}}{R^{\alpha + 1} (\alpha+1)!} g^{(\ell + \alpha)} \parenth{y + \frac{2 \pi k}{R}} \biggr|;
\end{align*}
\begin{align*}
    S_{2} & = \biggr|\sum_{k = -\infty}^{\infty} \biggr(\int_{y + \frac{2\pi k}{R}}^{y + \frac{2\pi(k + 1)}{R}}\frac{\parenth{x - y - \frac{2\pi k}{R}}^{K - \ell}}{(K - \ell - 1)!} \\
    &  \hspace{6 em} \times \int_{0}^{1} (1 - \xi)^{K - \ell - 1} g^{(K)} \parenth{(1 - \xi) \parenth{y + \frac{2 \pi k}{R}} + \xi x } d \xi \biggr) dx \biggr| \\
    & \leq \sum_{k = -\infty}^{\infty} \int_{y + \frac{2\pi k}{R}}^{y + \frac{2\pi(k + 1)}{R}} \int_{0}^{1} \frac{\parenth{x - y - \frac{2\pi k}{R}}^{K - \ell}}{(K - \ell - 1)!} (1 - \xi)^{K - \ell - 1} \sup_{t \in \brackets{y + \frac{2\pi k}{R}, y + \frac{2 \pi (k + 1)}{R}}} \abss{g^{(K)}(t)} d\xi dx \\
    & = \sum_{k = -\infty}^{\infty}(K - \ell) \frac{(2\pi)^{K - \ell + 1}}{R^{K - \ell + 1} (K - \ell+1)!} \sup_{t \in \brackets{y + \frac{2\pi k}{R}, y + \frac{2 \pi (k + 1)}{R}}} \abss{g^{(K)}(t)}.
\end{align*}
An application of the triangle inequality and the hypothesis of the induction argument shows that
\begin{align*}
    S_{1} & \leq \sum_{\alpha \leq K - 1 - \ell} \frac{(2\pi)^{\alpha + 1}}{R^{\alpha + 1} (\alpha+1)!} \abss{\sum_{k = -\infty}^{\infty} g^{(\ell + \alpha)} \parenth{y + \frac{2 \pi k}{R}}} \leq \frac{C_{2}}{R^{K - \ell}}
\end{align*}
where $C_{2}$ is some universal constant.

Similar to the setting $\ell = K - 1$, an application of the Riemann sum approximation theorem leads to
\begin{align*}
	S_{2} & \leq (K - \ell) \frac{(2\pi)^{K - \ell + 1}}{R^{K - \ell + 1} (K - \ell+1)!} \sum_{k = -\infty}^{\infty}  \sup_{t \in \brackets{y + \frac{2\pi k}{R}, y + \frac{2 \pi (k + 1)}{R}}} \abss{g^{(K)}(t)} \leq \frac{C_{3}}{R^{K - \ell}},
\end{align*}
when $R$ is sufficiently large where $C_{3}$ is some universal constant. Putting all the above results together, as long as $R \geq C$ we have
\begin{align*}
    \biggr|\sum_{k = -\infty}^{\infty} g^{(\ell)} \parenth{y + \frac{2 \pi k}{R}} \frac{(2\pi)}{R} - \int_{\mathbb{R}} g^{(\ell)}(x) dx \biggr| \leq \frac{\bar{C}}{R^{K - \ell}}
\end{align*}
for some universal constant $\bar{C}$. As $\int_{\mathbb{R}} g^{(\ell)}(x) dx = 0$,
the above inequality leads to the conclusion of equation~\eqref{eq:inequality_first} for $1 \leq \ell \leq K - 1$. As a consequence, we obtain the conclusion of equation~\eqref{eq:inequality_first} for all $\ell \in \{0, 1, \ldots, K - 1\}$. 

Given equation~\eqref{eq:inequality_first}, an application of the triangle inequality leads to
\begin{align}
    \abss{A_{\ell}} \leq \frac{1}{R^{\ell}} \abss{\int_{[0, 2\pi)} \frac{t^{\ell} \sin(t)}{\ell!} dt} \abss{ \sum_{k = -\infty}^{\infty} g^{(\ell)} \parenth{y + \frac{2 \pi k}{R}}} \leq \frac{\bar{C}}{R^{K - 1}} \label{eq:bound_Al}
\end{align} 
for all $0 \leq \ell \leq K - 1$. 

We now find a bound for $\abss{A_{K}}$. A direct application of the triangle inequality leads to the following bound of $\abss{A_{K}}$:
\begin{align*}
	\abss{A_{K}} \leq \int_{[0, 2\pi)} \frac{\abss{\sin(t) t^{K}}}{R^{K}(K - 1)!} dt \parenth{\sum_{k = -\infty}^{\infty} \int_{0}^{1} (1 - \xi)^{K - 1} \abss{g^{(K)} \parenth{y + \frac{2 \pi k}{R} + \frac{\xi t}{R}}} d \xi}.
\end{align*}
For any $\xi \in [0,1]$ and $t \in [0, 2\pi)$, we have
\begin{align*}
    \abss{g^{(K)} \parenth{y + \frac{2 \pi k}{R} + \frac{\xi t}{R}}} \leq \sup_{x \in \brackets{y + \frac{2\pi k}{R}, y + \frac{2\pi (k + 1)}{R}}} \abss{g^{(K)}(x)}.
\end{align*}
Putting the above inequalities together, we find that
\begin{align*}
    \abss{A_{K}} & \leq \int_{[0, 2\pi)} \frac{\abss{\sin(t) t^{K}}}{R^{K} K!} dt \parenth{\sum_{k = -\infty}^{\infty} \sup_{x \in \brackets{y + \frac{2\pi k}{R}, y + \frac{2\pi (k + 1)}{R}}} \abss{g^{(K)}(x)}}.
\end{align*}
From the Riemann sum approximation theorem, we obtain
\begin{align*}
    \frac{(2\pi)}{R^{1 - K}} \abss{A_{K}} \leq C_{1} \int_{[0, 2\pi)} \frac{\abss{\sin(t) t^{K}}}{R^{K} K!} dt,
\end{align*}
where $C_{1}$ is some universal constant. Collecting the above results, we conclude that 
\begin{align}
    |A_{K}| \leq \frac{\bar{C}}{R^{K - 1}} \label{eq:bound_AK}
\end{align}
for some constant $\bar{C}$. 
Putting the bounds~\eqref{eq:bound_Al} and~\eqref{eq:bound_AK} into equation~\eqref{eq:key_inequality_Fourier}, we obtain the conclusion of the theorem when $d = 1$.

\vspace{0.5 em}
We now provide the proof of Theorem~\ref{theorem:approximation_Fourier} for general dimension $d$.

\noindent
When $m(x) = \sum_{j = 1}^{d} m_{j}(x_{j})$ for any $x = (x_{1}, \ldots, x_{d})$: From the definition of $m_{R}(y)$ in equation~\eqref{eq:Fourier_approx}, we have
\begin{align*}
    \abss{m_{R}(y) - m(y)} & = \abss{\frac{1}{\pi^{d}} \int_{\mathbb{R}^{d}} \prod_{j = 1}^{d} \frac{\sin(R(y_j-x_j))}{(y_{j} - x_{j})} \parenth{m(x) - m(y)} dx} \\
    & = \abss{\frac{1}{\pi^{d}} \int_{\mathbb{R}^{d}} \prod_{j = 1}^{d} \frac{\sin(R(y_j-x_j))}{(y_{j} - x_{j})} \parenth{\sum_{j = 1}^{d} m_{j}(x_{j}) - \sum_{j = 1}^{d} m_{j}(y_{j})} dx} \\
    & \leq \sum_{j = 1}^{d} \abss{\frac{1}{\pi} \int_{\mathbb{R}} \frac{\sin(R(y_j-x_j))}{(y_{j} - x_{j})} \parenth{m_{j}(x_{j}) - m_{j}(y_{j})} dx_{j}}.
\end{align*}
Since $m_{j} \in \mathcal{T}^{K_{j}}(\mathbb{R})$ for $1 \leq j \leq d$, an application of the result of Theorem~\ref{theorem:approximation_Fourier} when $d = 1$ leads to 
\begin{align*}
	\abss{\frac{1}{\pi} \int_{\mathbb{R}} \frac{\sin(R(y_j-x_j))}{(y_{j} - x_{j})} \parenth{m_{j}(x_{j}) - m_{j}(y_{j})} dx_{j}} \leq \frac{C_{j}}{R^{K_{j}}}
\end{align*}
where $C_{j}$ are universal constants. Putting the above results together, we obtain the conclusion of the theorem when $m$ is the summation of the functions $m_{1}, m_{2}, \ldots, m_{d}$.

\vspace{0.5 em}
\noindent
When $m(x) = \prod_{j = 1}^{d} m_{j}(x_{j})$ for any $x = (x_{1}, \ldots, x_{d})$: Similar to the argument when $m$ is the summation of $m_{1}, m_{2}, \ldots, m_{d}$, we have
\begin{align*}
    \abss{m_{R}(y) - m(y)} & = \abss{\frac{1}{\pi^{d}} \int_{\mathbb{R}^{d}} \prod_{j = 1}^{d} \frac{\sin(R(y_j-x_j))}{(y_{j} - x_{j})} \parenth{\prod_{j = 1}^{d} m_{j}(x_{j}) - \prod_{j = 1}^{d} m_{j}(y_{j})} dx} \\
    & = \biggr|\frac{1}{\pi^{d}} \int_{\mathbb{R}^{d}} \prod_{j = 1}^{d} \frac{\sin(R(y_j-x_j))}{(y_{j} - x_{j})} \biggr(\sum_{\ell = 0}^{d - 1} \prod_{j = 1}^{\ell} m_{j}(y_{j}) \prod_{j = \ell + 1}^{d} m_{j}(x_{j}) \\
    & \hspace{14 em} - \prod_{j = 1}^{\ell + 1} m_{j}(y_{j}) \prod_{j = \ell + 2}^{d} m_{j}(x_{j})\biggr) dx\biggr| \\
    & \leq \sum_{\ell = 0}^{d - 1} \biggr|\frac{1}{\pi^{d}} \int_{\mathbb{R}^{d}} \prod_{j = 1}^{d} \frac{\sin(R(y_j-x_j))}{(y_{j} - x_{j})} \biggr(\prod_{j = 1}^{\ell} m_{j}(y_{j}) \prod_{j = \ell + 1}^{d} m_{j}(x_{j}) \\
    & \hspace{14 em} - \prod_{j = 1}^{\ell + 1} m_{j}(y_{j}) \prod_{j = \ell + 2}^{d} m_{j}(x_{j})\biggr) dx \biggr| \\
    & \leq C \sum_{\ell = 0}^{d - 1} \abss{\frac{1}{\pi} \int_{\mathbb{R}} \frac{\sin(R(y_{\ell + 1}-x_{\ell + 1}))}{(y_{\ell + 1} - x_{\ell + 1})} \parenth{m_{\ell + 1}(x_{\ell + 1}) - m_{\ell + 1}(y_{\ell + 1})} dx_{\ell + 1}}
\end{align*}
where $C$ is some universal constant. Using the above bound and the result in one dimension of Theorem~\ref{theorem:approximation_Fourier} for $m_{1}, m_{2}, \ldots, m_{d}$, we obtain the conclusion of the theorem when $m$ is the product of these functions. 

\bibliographystyle{plainnat}
\bibliography{Nhat}

\begin{thebibliography}{21}
\providecommand{\natexlab}[1]{#1}
\providecommand{\url}[1]{\texttt{#1}}
\expandafter\ifx\csname urlstyle\endcsname\relax
  \providecommand{\doi}[1]{doi: #1}\else
  \providecommand{\doi}{doi: \begingroup \urlstyle{rm}\Url}\fi

\bibitem[Chacon and Duong(2018)]{Chacon18}
J.E. Chacon and T.~Duong.
\newblock \emph{Multivariate Kernel Smoothing and its Applications}.
\newblock CRC Press, 2018.

\bibitem[Chen and Huang(2007)]{Chen2007}
S.X. Chen and T.M. Huang.
\newblock Nonparametric estimation of copula functions for dependence
  modelling.
\newblock \emph{Canadian Journal of Statistics}, 35:\penalty0 265--282, 2007.

\bibitem[Daouia et~al.(2013)Daouia, Gardes, and Girard]{Daouia2013}
A.~Daouia, L.~Gardes, and S.~Girard.
\newblock On kernel smoothing for extremal quantile regression.
\newblock \emph{Bernoulli}, pages 2557--2589, 2013.

\bibitem[Davis(1975)]{Davis75}
K.B. Davis.
\newblock Mean square error properties of density estimates.
\newblock \emph{Annals of Statistics}, 3:\penalty0 1025--1030, 1975.

\bibitem[Geenens et~al.(2017)Geenens, Charpentier, and
  Paindaveine]{Geenens2017}
G.~Geenens, A.~Charpentier, and D.~Paindaveine.
\newblock Probit transformation for nonparametric kernel estimation of the
  copula density.
\newblock \emph{Bernoulli}, 23:\penalty0 1848--1873, 2017.

\bibitem[Hall et~al.(1999)Hall, Wolff, and Tao]{Hall99}
P.~Hall, R.C. Wolff, and Q.~Tao.
\newblock Methods for estimating a conditional distribution function.
\newblock \emph{Journal of the American Statistical Association}, 94:\penalty0
  154--163, 1999.

\bibitem[Hayfield and Racine(2008)]{Hayfield08}
T.~Hayfield and J.S. Racine.
\newblock Nonparametric econometrics: The np package.
\newblock \emph{Journal of Statistical Software}, 27, 2008.

\bibitem[Ho and Walker(2021)]{Ho21}
N.~Ho and S.G. Walker.
\newblock Multivariate smoothing via the {F}ourier integral theorem and
  {F}ourier kernel.
\newblock \emph{Arxiv preprint Arxiv:2012.14482}, 2021.

\bibitem[Jin and Shao(1999)]{Jin99}
J.~Jin and Y.~Shao.
\newblock On kernel estimation of a multivariate distribution function.
\newblock \emph{Statistics and Probability Letters}, 41:\penalty0 163--168,
  1999.

\bibitem[Koenker(2005)]{Koenker2005}
R.~Koenker.
\newblock \emph{Quantile Regression}.
\newblock Cambridge University Press, 2005.

\bibitem[Koenker and Bassett(1978)]{Koenker1978}
R.~Koenker and G.~Bassett.
\newblock Regression quantiles.
\newblock \emph{Economatrica}, 46:\penalty0 33--50, 1978.

\bibitem[Liu and Wu(2011)]{Liu2011}
Y.~Liu and Y.~Wu.
\newblock Simultaneous multiple non--crossing quantile regression.
\newblock \emph{Journal of Nonparametric Statistics}, 23:\penalty0 415--437,
  2011.

\bibitem[Panaretos and Konis(2012)]{Panaretos12}
V.M. Panaretos and K.~Konis.
\newblock Nonparametric construction of multivariate kernels.
\newblock \emph{Journal of the American Statistical Association}, 107:\penalty0
  1085--1095, 2012.

\bibitem[Parzen(1962)]{Parzen62}
E.~Parzen.
\newblock On estimation of a probability density function and mode.
\newblock \emph{Annals of Mathematical Statistics}, 33:\penalty0 1065--1076,
  1962.

\bibitem[Sestelo et~al.(2017)Sestelo, Meira-Machado, Villanueva, and
  Roca-Pardinas]{Sestelo17}
M.~Sestelo, L.~Meira-Machado, N.M. Villanueva, and J.~Roca-Pardinas.
\newblock npregfast: An r package for nonparametric estimation and inference in
  life sciences.
\newblock \emph{Journal of Statistical Software}, 82, 2017.

\bibitem[Staniswalis et~al.(1993)Staniswalis, Messer, and Finston]{Stanis93}
J.G. Staniswalis, K.~Messer, and D.R. Finston.
\newblock Kernel estimators for multivariate regression.
\newblock \emph{Journal of Nonparametric Statistics}, 3:\penalty0 103--121,
  1993.

\bibitem[Takeuchi et~al.(2005)Takeuchi, Le, Sears, and Smola]{Takeuchi2005}
I.~Takeuchi, Q.V. Le, T.~Sears, and A.J. Smola.
\newblock Nonparametric quantile regression.
\newblock \emph{Journal of Machine Learning Research}, 7, 2005.

\bibitem[Veraverbeke et~al.(2014)Veraverbeke, Gijbels, and Omelka]{Gijbels14}
N.~Veraverbeke, I.~Gijbels, and M.~Omelka.
\newblock Preadjusted non--parametric estimation of a conditional distribution
  function.
\newblock \emph{Journal of the Royal Statistical Society, Series B},
  76:\penalty0 399--438, 2014.

\bibitem[Wand(1992)]{Wand92}
M.P. Wand.
\newblock Error analysis for general multivariate kernel estimators.
\newblock \emph{Journal of Nonparametric Statistics}, 2:\penalty0 1--15, 1992.

\bibitem[Wand and Jones(1993)]{WJones93}
M.P. Wand and M.C. Jones.
\newblock Comparison of smoothing parameterizations in bivariate kernel density
  estimation.
\newblock \emph{Journal of the American Statistical Association}, 88:\penalty0
  520--528, 1993.

\bibitem[Yu and Jones(1998)]{Yu1998}
K.~Yu and M.C. Jones.
\newblock Local linear quantile regression.
\newblock \emph{Journal of the American Statistical Association}, 93:\penalty0
  228--237, 1998.

\end{thebibliography}
\end{document}